\documentclass{ws-mpla}
\usepackage[super]{cite}
\usepackage{mathrsfs}
\usepackage{graphicx,bm,amsmath,amssymb,latexsym,psfrag,epsfig,euscript, multirow,color,verbatim,cancel}
\allowdisplaybreaks

\def\be{\begin{equation}}
\def\ee{\end{equation}}
\def\bea{\begin{eqnarray}}
\def\eea{\end{eqnarray}}
\def\bit{\begin{itemlist}}
\def\eit{\end{itemlist}}
\newcommand{\Tf}{T_{\rm f}}
\newcommand{\mpl}{M_{\rm pl}}
\newcommand{\tb}{\tilde{B}}

\begin{document}

%{Yanou Cui}
%{Instructions for Typing Manuscripts (Paper's Title)}

%%%%%%%%%%%%%%%%%%%%% Publisher's Area please ignore %%%%%%%%%%%%%%
\catchline{}{}{}{}{}
%%%%%%%%%%%%%%%%%%%%%%%%%%%%%%%%%%%%%%%%%%%%%%%%%%%%%%%%%%%%%%%%%%%

\title{A REVIEW OF WIMP BARYOGENESIS MECHANISMS }

\author{\footnotesize YANOU CUI}

\address{Perimeter Institute for Theoretical Physics, 31 Caroline St. North,\\
Waterloo, ON N2L 2Y5,
Canada\\
Maryland Center for Fundamental Physics, Department of
Physics, University of Maryland, College Park, MD 20742, USA\\
ycui@perimeterinstitute.ca}

\maketitle

%\pub{Received (Day Month Year)}{Revised (Day Month Year)}

\begin{abstract}
It was recently proposed that weakly interacting massive particles (WIMP) may provide new ways of generating the observed baryon asymmetry in the early universe, as well as addressing the cosmic coincidence between dark matter and baryon abundances. This suggests a new possible connection between weak scale new particle physics and modern cosmology. This review summarizes the general ideas and simple model examples of the two recently proposed WIMP baryogenesis mechanisms: baryogenesis from WIMP dark matter annihilation during thermal freezeout, and baryogenesis from metastable WIMP decay after thermal freezeout. This letter also discusses the interesting phenomenology of these models, in particular the experimental signals that can be probed in the intensity frontier experiments and the Large Hadron Collider (LHC) experiments.

\end{abstract}

\ccode{PACS Nos.: 95.30Cq, 95.35+d, 98.80Cq}

\section{Introduction}

\indent
    Over the past two decades, we have seen the beginning and rapid development of the new era of precision cosmology. The combinations of a variety of experiments have provided us unprecedented knowledge about energy composition in our universe. In particular, now we know very well that the visible atomic (baryonic) matter that composes ourselves takes up $\approx 4\%$ of the cosmic energy budget, while $\approx 26\%$ of the cosmic energy or $85\%$ of total matter density in the universe is composed of invisible dark matter (DM) \cite{Ade:2015xua}. Such new knowledge in fact leaves us with more pressing puzzles to resolve, and has motivated tremendous efforts in related research, as these observations cannot be explained with the known particles and interactions within the Standard Model (SM) of modern particle physics. None of the SM particles can be an eligible DM candidate, while the prediction for baryon abundance based on the ingredients in the SM is well below the observed value. In addition, the comparability of the amounts of DM and baryonic matter today appears to be a puzzling cosmic ``coincidence", considering the feeble interaction between DM and visible matter in the current time. These important questions imposed by modern cosmological observations-- dark matter, the origin of baryonic matter (so-called ``baryogenesis"), and the above cosmic coincidence problem -- strongly motivate new physics beyond the framework of the SM.\\

On the other hand, the Standard Model of particle physics has its own frontier and challenges at the ``weak scale" energies. The name ``weak scale" originates from weak interaction in the SM, which enables important physics processes such as radiative $\beta$-decay of neutrons. The strength of weak interaction is characterized by the Fermi coupling constant $G_F=1.166 ~\rm GeV^{-2}$. This corresponds to a mediator particle ($W$-boson, in the case of $\beta$-decay) mass of $\sim100$ GeV. The weak scale, being $O(100)$ times of the proton mass, is the energy frontier of modern particle physics. The most massive known particles in the SM are of weak scale mass, such as $W$, $Z$ gauge bosons, top quark and the recently discovered Higgs boson. There are compelling reasons to expect new weak scale particles beyond the SM, in particular motivated by the naturalness or hierarchy problem at weak scale related to the radiative stability of the Higgs mass. Starting 2008, the Large Hadron Collider (LHC), the largest and most powerful particle accelerator ever, was dedicated to searching for new particles that can emerge at weak scale energies, with the Higgs discovery as its first milestone. 

A natural curiosity is: could some of the new weak scale particles to be discovered also shed light on the prominent puzzles in modern cosmology as we summarized earlier on? Could they be dark matter, and/or trigger the generation of baryon abundance? Such potential connection or interface between modern particle physics and modern cosmology is rather appealing, and has drawn a great amount of research efforts. A lot of such efforts have been focusing on the so-called ``WIMP''-type of particle as DM candidate. ``WIMP'' stands for  weakly interacting massive particle, which by definition has a weak scale mass and a weak scale interaction strength of $\sim G_F$, thus it is a naturally expected type of particle that can appear in the new energy frontier which we are probing.  There are concrete candidates of WIMP DM from motivated theory frameworks aimed at solving the weak scale hierarchy problem, e.g. Lightest Supersymmetric Particle (LSP) in supersymmetry (SUSY), Kaluza-Klein (KK) state of photon as in extra-dimension model. WIMP DM is a compelling candidate for DM as the scenario naturally predicts the observed relic abundance based on the thermal history, in particular thermal freeze-out of the WIMP. In Section.\ref{sec:warmup} we will briefly review this WIMP ``miracle'' DM scenario. 

Recently there have been proposals \cite{Cui:2011ab, McDonald:2011zza, Davidson:2012fn, Cui:2012jh} demonstrating a new direction of possible connection between modern particle physics and modern cosmology: baryogenesis may be triggered by a WIMP-type of particle, and in such a scenario the coincidence between DM and baryon abundance can also be addressed. The two major proposals are baryogenesis from WIMP annihilation (``WIMPy baryogenesis") and baryogenesis from metastable WIMP decay. These proposals are based on the realization that thermal freeze-out of WIMP naturally provides the crucial out-of-equilibrium condition for baryogenesis. The embedding of these general ideas in motivated theory framework addressing the weak scale hierarchy problem has also been studied. Although the original motivation for these proposals was to provide a novel of way addressing the baryon-DM coincidence problem while maintaining WIMP DM miracle, they can be seen as new low scale baryogenesis mechanisms independent of the specifics of DM. In contrast to many of the conventional baryogenesis or leptogenesis models which involve new physics well above weak scale, these new models contain new particles with relatively low mass, and thus can be within the reach of the current or near future particle collider experiments and intensity frontier experiments. This review is dedicated to giving an overview of the basic theoretical ideas and example models of the two types of WIMP baryogenesis mechanisms, as well as their phenomenology implications for a variety of particle physics experiments.

\section{Warm-up: Thermal History of WIMP and Baryogenesis Basics}\label{sec:warmup}
\subsection{Thermal History of a WIMP-type of Particle}
\begin{figure}
\begin{center}
\includegraphics[width=8cm]{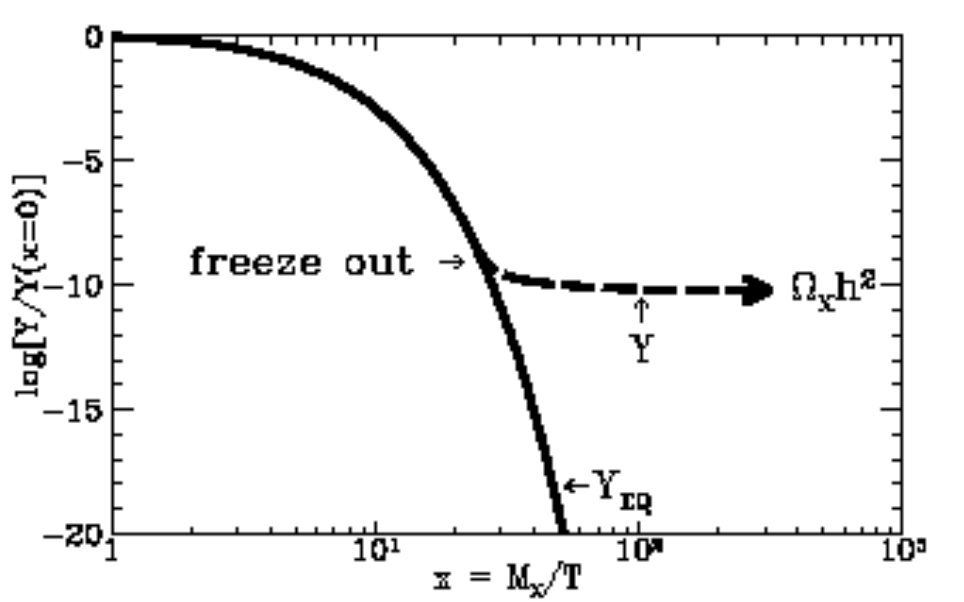}
\caption{Thermal evolution of the co-moving density of a stable WIMP particle.
}\label{fig:DM_evol}
\end{center}
\end{figure}

Now let us first review the thermal history of a stable WIMP particle $\chi$ which can be a candidate for DM. As we will discuss later, this picture can also apply to more generic WIMP that can be cosmologically unstable. One of the most appealing features of WIMP DM is that its relic abundance is simply set by the thermodynamics of the system where DM interacts with other particles in the expanding universe, and is insensitive to cosmic initial conditions which bear large uncertainty. In the very early time of our universe, shortly after the hot big bang, WIMP DM maintains in thermal equilibrium with the SM particles via rapid 2-to-2 annihilation (for variations such as annihilating into dark sector particles, see for instance, \cite{Pospelov:2007mp, Belanger:2011ww, Agashe:2014yua, Chacko:2015noa}). This annihilation rate is approximately $\Gamma_{\rm ann}\sim n_\chi^{\rm eq}\langle\sigma_{\rm ann} v\rangle$, where $n_\chi^{\rm eq}$ is the number density of $\chi$ following the equilibrium distribution. As the universe expands, the cosmic temperature $T$ drops below the $\chi$ mass, $m_\chi$, when $n_\chi^{\rm eq}$ starts to bear a Boltzmann suppression factor. At an even later time, $\Gamma_{\rm ann}$ further decreases and falls below the Hubble expansion rate $H$, when $\chi$ can no longer stay in equilibrium. This stage of $\chi$ departure from thermal equilibrium, around the time when $\Gamma_{\rm ann}\sim H$ is the so-called ``thermal freeze-out''. The evolution of $\chi$ number density can be obtained by solving the following Boltzmann equation:
 \be
\frac{\mathrm{d}n_\chi}{\mathrm{d}t}+3Hn_\chi=-\langle \sigma v \rangle\left(n_\chi^2-(n_\chi^{\rm eq})^2  \right).
\label{eq:boltzmann}
 \ee
In the radiation dominated universe, we can rewrite the above eq. in terms of variable $x\equiv m_\chi/T$ and co-moving density $Y_\chi\equiv n_\chi/s$ where $s$ is the entropy density of the universe: 
\be
\frac{\mathrm{d}Y_\chi}{\mathrm{d}x} = - \frac{\langle \sigma  v \rangle}{H x} s \left( Y_\chi^2 - (Y_{\chi}^\text{eq})^2 \right),
\label{eq:dYdx_cano}
 \ee
where $g$ counts the internal degrees of freedom of $\chi$, $g_*$ counts the total number of relativistic degrees of freedom in the thermal bath, the co-moving density $Y_\chi$ solution approaches its late-time asymptotic value after thermal freezeout at temperature $T_f$. The accurate solution can be obtained numerically. The analytic approximate solution at leading order is as follows:
\bea
T_f&\simeq& m_\chi\left[\ln\left(0.038({g}/{g_*^{1/2}})m_\chi\mpl\langle\sigma_{\rm A} v\rangle\right)\right]^{-1},\label{wimpfo}\\
Y_\chi(\Tf)&\simeq& 3.8\frac{g_*^{1/2}}{g_{*s}}\frac{m_\chi}{\Tf}\left(m_\chi\mpl\langle\sigma_{\rm A} v\rangle\right)^{-1}. \label{chiinitial}
\eea
The relic abundance of $\chi$, as a result of the thermal freeze-out is:
\bea
     \Omega_{\chi}&=&\frac{m_\chi Y_\chi(\Tf)s_0}{\rho_0}\simeq0.1\frac{\alpha_{\rm weak}^2/(\rm TeV)^2}{\langle\sigma_{\rm A} v\rangle}\\
    & \simeq& 0.1\left(\frac{G_F}{G_\chi}  \right)^2\left(\frac{m_{\rm weak}}{m_\chi}  \right)^2,
   \label{omegawimp}
\eea
where $G_\chi\sim g^2_\chi/m_{\rm med}^2$ ($m_{\rm med}$ being the mass of mediator for DM annihilation) characterizes the interaction strength of the WIMP $\chi$ annihilation, is analogous to Fermi constant $G_F$. We see that with weak scale mass and weak scale interaction, the prediction for the relic abundance of a stable WIMP readily agrees with the observed value of DM abundance. This is the so-called ``WIMP miracle''. Fig.\ref{fig:DM_evol} gives an intuitive illustration of the evolution of $Y_\chi$.

Before moving on, here are a couple of key points to take note on as they are essential for our later discussion on WMP triggered baryogenesis:
\bit
\item  Although the WIMP miracle is rather impressive, it is not a precise prediction down to order one level. As we can see from eq.\ref{omegawimp}, varying the couplings or masses by O(1) can lead to a few orders of magnitudes variation in the prediction for relic abundance. For instance, in the Minimal Supersymmetric Standard Model (MSSM) with conserved R-parity symmetry, neutralinos (bino, wino, higgsino) are all WIMP particles, but can have very different relic abundances due to different couplings and masses.
\item  The departure from the equilibrium distribution at the freeze-out stage is essential for WIMP DM to have sizeable relic abundance today.  As can be seen clearly from Fig.\ref{fig:DM_evol}, without the departure from equilibrium, $\chi$ abundance would follow the Boltzmann distribution which is exponentially suppressed at late time.
\eit

\subsection{A Mini-Review of the Conditions for Baryogenesis}\label{sec:BGreview}
Now let's switch gear to review some basics for generating a baryon asymmetry in the early universe (baryogenesis). Although it is plausible that the matter-anti-matter asymmetry results from certain initial condition at the hot Big Bang, this scenario is disfavoured by the recently established paradigm of cosmic inflation which implies that any initial asymmetry would be diluted away during inflation, when the size of the universe exponentially grows. Therefore a much more likely scenario is that the baryon asymmetry is generated dynamically after inflation. In 1967, in his seminal paper \cite{Sakharov:1967dj}, Sakharov pointed out three conditions for generating a baryon asymmetry dynamically, assuming exact CPT symmetry (for exceptions, see e.g.\cite{Cohen:1988kt}):
\bit
\item Baryon number (B) violation, or lepton number (L) violation before the shut-off of sphaleron process which converts L to B,
\item C and CP violation,
\item Departure from thermal equilibrium.
\eit
The first two conditions are more intuitive to understand. A quick way to explain the third condition is that: by CPT symmetry, baryon and anti-baryon always have exactly the same mass, which determines that their density in equilibrium always exactly equal to each other, and thus the net baryon number would be 0 in thermal equilibrium. A more detailed/rigorous discussion about these conditions can be found in \cite{Kolb:1979qa, Bernreuther:2002uj}. In a concrete particle physics model, the first two conditions can be arranged by proper interactions involving complex phases. The third condition is more non-trivial to realize. In the electroweak baryogenesis mechanism\cite{Cohen:1990it}, the out-of-equilibrium condition is realized by bubble collision and expansion, while in leptogenesis mechanism related to neutrino mass generation, it is realized by out-of-equilibrium decay of a massive particle. Here it is rather intriguing to note that departure from equilibrium is essential for both generating cosmic baryon abundance as well as WIMP DM relic abundance, as we just discussed earlier in this Section. We will get back to this observation very soon. \\

In addition to the above Sakharov conditions, there are two other more subtle considerations for a successful baryogenesis:

\bit
\item In association with process that generate a baryon asymmetry, there are often other B- or L-violating processes, the so-called washout processes, that could erase the generated asymmetry, or reduce the efficiency of baryogenesis. Such washout effect needs to be under control to ensure a sufficient baryon abundance today.
\item Weinberg-Nanopolous theorem \cite{Nanopoulos:1979gx} imposes another condition for a general class of baryogenesis models, e.g. those based on out-of-equilibrium decay of a massive particle. The theorem states that in order to create non-vanishing baryon asymmetry, there must be B-violating source at second-order (e.g. via interference loop process) in addition to B-violation at leading order (e.g. decay by tree-level process).
\eit

\subsection{The Inspiring Questions and Outline of New Ideas}

With the above essential ingredients of WIMP DM and general baryogenesis in mind, one could ask the following inspiring questions which led to the novel baryogenesis ideas that we will discuss:
\bit
\item As noted earlier, departure from equilibrium is crucial for establishing relic abundances of both WIMP DM and baryons. It is then curious to ask: could thermal freeze-out, the process by which a WIMP particle departs from equilibrium also provide the 3rd Sakharov condition for baryogenesis? 
\item If the above scenario can be realized, baryogenesis would relate to weak scale new particle physics. Then is there any potential of directly testing such baryogenesis mechanism at current-day experiments such as the LHC? If so, this would be in direct analogy to the appealing prospect of searching for WIMP DM, while is in contrast to most of existing baryogenesis mechanisms which are hard to directly test due to high mass or high temperature that is necessary. 
\item The possibility of baryogenesis triggered by WIMP freeze-out suggests that there could be a WIMP miracle prediction for baryon abundance, in analogy to that for DM. Assuming DM is indeed a WIMP particle, this shared WIMP miracle can provide a novel way to address the cosmic coincidence between DM and baryon abundance. It suggests that the solution to the coincidence problem can be compatible with WIMP DM, and it does not necessarily tie to asymmetric DM scenario.
\eit

In the following sections we will summarize the proposals and developments of two concrete new baryogenesis ideas that were motivated by these questions. In the first proposal, baryon asymmetry is produced by out-of-equilibrium annihilation of a stable WIMP DM. In the second proposal, we consider a general WIMP particle which in fact decays after the thermal annihilation freezes out and produce baryon asymmetry. As we will see, the latter proposal has more robust prediction for baryon abundance, and distinct signature in collider experiments such as the LHC, while both ideas have theoretical appeals along with rich phenomenology in various experiments.

\section{Baryogenesis from WIMP Annihilation}
\subsection{General Idea}
The first type of mechanism is the so called ``WIMPy Baryogenesis'', proposed in 2011 \cite{Cui:2011ab}. Here we consider the possibility that WIMP DM annihilation violates baryon or lepton number, as well as C and CP symmetries. These together with the out-of-equilibrium condition naturally provided by WIMP freeze-out make it possible to satisfy all the three Sakharov conditions. Therefore in this scenario the relic abundances of baryons and WIMP DM are simultaneously generated during the epoch of WIMP freeze-out, and this provides a novel way to address the $\Omega_B-\Omega_{\rm DM}$ coincidence, while preserving the WIMP miracle prediction for the absolute amount of $\Omega_{\rm DM}$.  Nonetheless, as we will show, the prediction for baryon asymmetry is sensitive to the model parameters, which is a slight drawback of this otherwise neat scenario.\\

As reviewed earlier in Section.\ref{sec:BGreview}, the asymmetry-generating baryogenesis process is often accompanied by washout processes which in general need to be suppressed, in order to ensure a good efficiency of baryogenesis. In the models we are considering, the two leading washout processes are the inverse annihilations of baryons into dark matter and baryon-to-antibaryon scattering. To facilitate our discussion, it is convenient to define the freeze-out of washout as the time when the rate of the washout process falls below the Hubble expansion rate, which is in analogy to the freeze-out of WIMP annihilation. Any asymmetry created before the washout freeze-out is rapidly damped away. That is, baryon asymmetry can start to accumulate efficiently through DM annihilation, only after washout processes freeze out. 
The final asymmetry is determined by the amount of WIMP DM density that remains when washout scatterings freeze out. Apparently after WIMP DM freeze-out, the annihilation is no longer efficient for generating a sizeable baryon asymmetry, therefore the freeze-out of the washout processes must occur prior to that of WIMP annihilation. Before presenting the quantitative analysis, we may preview our central results here: \emph{if washout processes freeze out  before WIMP freeze-out, then a large baryon asymmetry may accumulate, and its final value is proportional to the WIMP abundance at the time that washout becomes inefficient}.\\

The above condition of early freeze-out of washout processes impose constraints on the possibilities of viable models. At $T<m_{\rm DM}$, the inverse process of WIMP annihilation to light baryonic states is naturally suppressed by the Boltzmann factor $\sim e^{-m_{\rm DM}/T}$ on the thermal number density of the baryon fields. However, the baryon-to-anti-baryon washout may still persist till well after WIMP freeze-out, as the SM baryons are light, and thus Boltzmann suppression on washout processes only take effect at a much lower temperature. In general, without the cost of reducing the rate of WIMP annihilation, such washout effect can only be suppressed by introducing a heavy exotic baryon in the initial state, with a mass $\gtrsim m_{\rm DM}$. In this way, at $T<m_{\rm DM}$ the washout scatterings get a stronger Boltzmann suppression than the WIMP annihilation, and thus freeze out earlier than the WIMP DM. Notice that on the other hand, in order for the WIMP to exotic baryon annihilation to be kinematically allowed, the exotic baryon needs to be lighter than about twice of the WIMP DM mass. Putting together these conditions, we see that in order to have efficient baryogenesis, the mass relation $m_{\rm DM}\lesssim m_{\rm exotic\,baryon}\lesssim 2m_{\rm DM}$ needs to be satisfied. This mass relation restricts the viable model parameter space in a specific region, although it does not require serious fine-tuning. 

\begin{figure}
\begin{center}
\includegraphics[width=5.5cm]{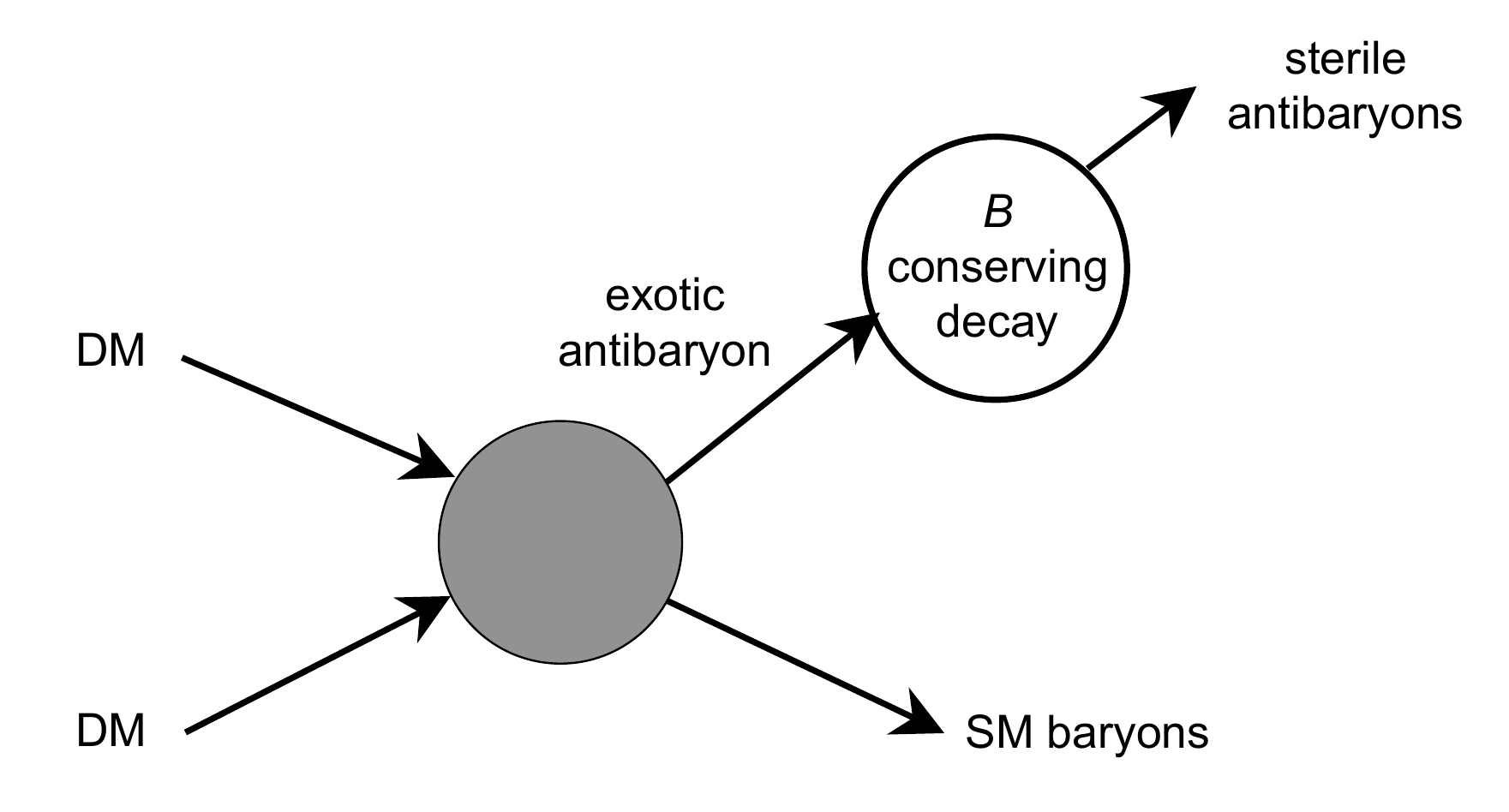}\hspace{1cm}
\includegraphics[width=5.5cm]{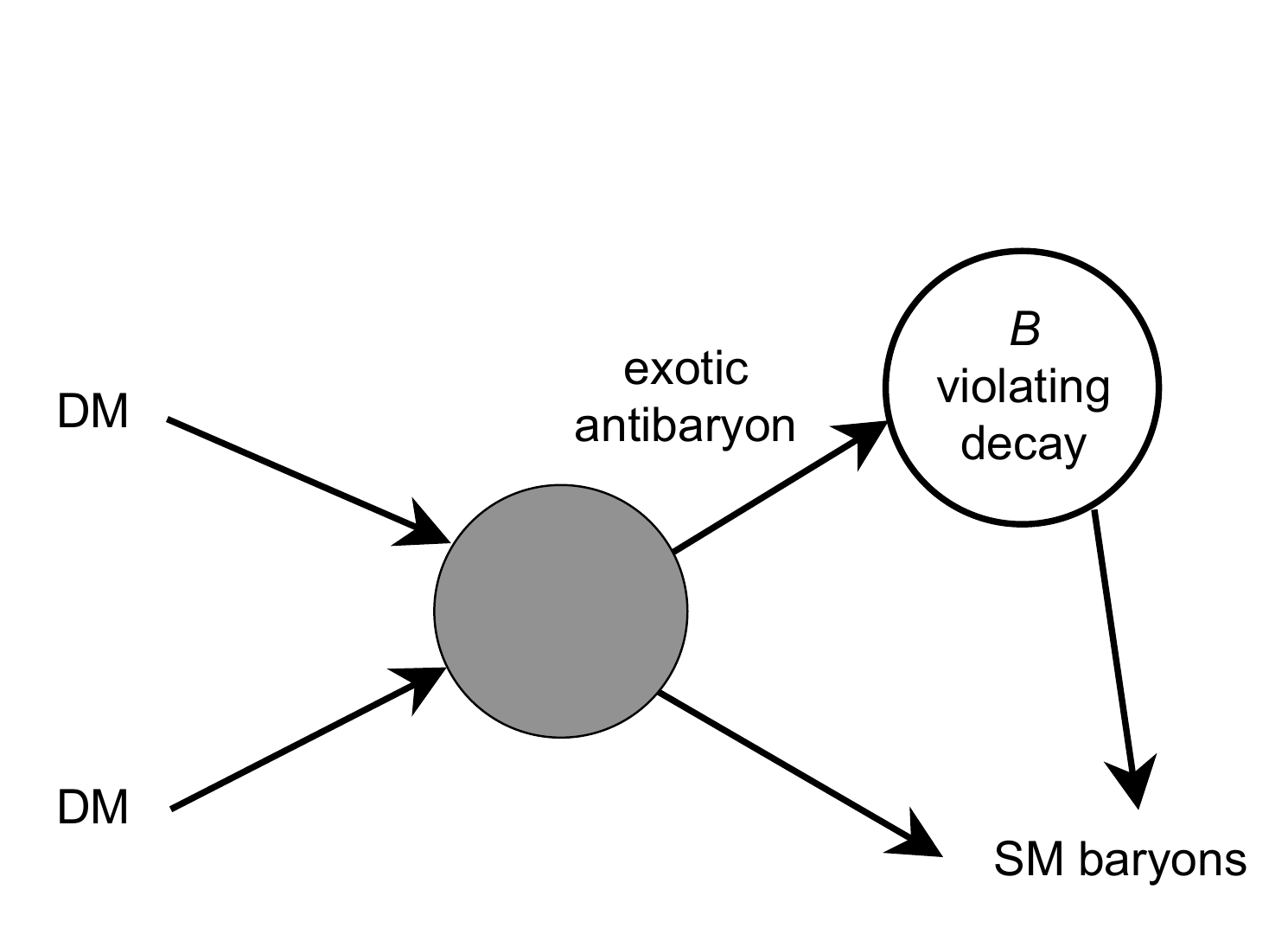}
\caption{Schematic diagrams showing the evolution of the asymmetry created by dark matter annihilation. {\bf (left)} Model where asymmetry created in exotic antibaryons is sequestered in a sterile sector through baryon-number-conserving decays. {\bf (right)} Model where asymmetry created in exotic antibaryons is converted into a Standard Model baryon asymmetry through baryon-number-violating decays.
}\label{fig:flow}
\end{center}
\end{figure}

The exotic baryon that freezes out before the WIMP would over-close the universe if it is cosmologically stable, and thus is required to decay.
 In Fig.\ref{fig:flow} we diagrammatically illustrate two general possibilities of this scenario according to the decay patterns of the exotic baryon (or lepton): DM annihilates to exotic baryon that decays though B-conserving interaction and finally store its B-number in very light sterile states (so that today there is a net baryon number in the SM sector), or DM annihilates to exotic baryon that decays through B-violating interaction into SM baryons.

\subsection{General formulation}

Now starting with Boltzmann equations, we will review the essential formulation and quantitative analysis for the evolution of WIMP DM and baryon asymmetry in this new mechanism. More details can be found in the original paper \cite{Cui:2011ab}. To a good approximation, the Boltzmann equation for the evolution of co-moving density of WIMP DM $X$, $Y_X$, is:
\be
\label{eq:YXgeneral}\frac{dY_X}{dx} =-\frac{2s(x)}{x\,H(x)}\,\langle\sigma_{\rm ann}v\rangle\left[Y_X^2-(Y_X^{\rm eq})^2\right],
\ee
where we neglect a back-reaction term $\epsilon\,s(x)\,\langle\sigma_{\rm ann}v\rangle\,Y_{\Delta B}(Y_X^{\rm eq})^2/(2Y_\gamma\,x\,H(x))$, with $Y_{\Delta B}$ being the comoving density of the net baryon number. $\langle\sigma_{\rm ann}v\rangle$ is the thermal averaged cross-section of WIMP annihilating to baryons and anti-baryons. $\epsilon<1$ is the CP asymmetry factor which it indicates the net baryon number produced per DM annihilation. As we will show in concrete model examples, the CP asymmetry $\epsilon$ typically comes from the interference between tree-level and loop diagram, and thus is generally estimated to be $\epsilon\sim10^{-2}$. This modification term is negligible when $Y_{\Delta B}/Y_\gamma\ll 1$, as is true in our universe ($\sim10^{-10}$). Apparently, with such simplification, $Y_{\Delta B}$ effectively decouples from the evolution of $Y_X$, and eq.\ref{eq:YXgeneral} restores the familiar Boltzmann equation for canonical WIMP DM eq.\ref{eq:dYdx_cano}. This makes it easier to get an approximate analytic solution for $Y_{\Delta B}$. \\

Here it is important to emphasize that, equilibrium distribution $Y_X=Y_X^{\rm eq}$ is only an exact solution of eq.\ref{eq:YXgeneral} when DM is relativistic, i.e. when $x\gg1$ and $Y_X$ is constant over $x$. As soon as $X$ become non-relativistic, i.e. starting from $x\sim1$, $Y_X^{\rm eq}$ no longer satisfies eq.\ref{eq:YXgeneral}. This tells that the departure from equilibrium starts as early as $x\sim1$, when the net baryon number can potentially start to be generated according to the 3rd Sakharov condition, although $Y_X$ still loosely track its equilibrium distribution. The conventional WIMP freeze-out, typically occurs at $x_F\sim 20-30$, at this later time the departure from $Y_X^{\rm eq}$ becomes significant and $Y_X$ approaches its asymptotic value. \\

The Boltzmann equation for the evolution of baryon asymmetry is more involved:
\be
\frac{dY_{\Delta B}}{dx} = \frac{\epsilon\,s(x)}{x\,H(x)}\,\langle\sigma_{\rm ann}v\rangle\left[Y_X^2-(Y_X^{\rm eq})^2\right] - \frac{s(x)}{x\,H(x)}\,\langle\sigma_{\rm washout}v\rangle \frac{Y_{\Delta B}}{2Y_\gamma}\prod_i Y^{\rm eq}_i.\label{eq:YBgeneral}
\ee
The first piece in eq.\ref{eq:YBgeneral}, proportional to $\langle\sigma_{\rm ann}v\rangle$, is the asymmetry generating source term from $X$ annihilation. It is clear that the amount of generated asymmetry is set by the depletion of WIMP DM through its annihilation, scaled by the fraction factor $\epsilon$. The second piece, proportional to $\langle\sigma_{\rm washout}v\rangle$, is the washout term from the B-violating scatterings that can potentially damp the asymmetry generated by WIMP annihilation. We have assumed that all other particles except for DM are in equilibrium. $\prod_i Y^{\rm eq}_i$ is the product of the equilibrium densities of initial states of washout processes, including the heavy exotic baryon/lepton that is necessary for a viable model.\\

Eq.\ref{eq:YBgeneral} manifests the Sakharov out-of-equilibrium condition: $Y_{\Delta B}$ would remain zero when all field are in equilibrium, with initial condition $Y_{\Delta B}=0$. We also see that the washout term is proportional to $Y_{\Delta B}$ itself, which suggests an exponential damping effect. Combining eqs.(\ref{eq:YXgeneral}, \ref{eq:YBgeneral}), we can write the solution for $Y_{\Delta B}$ in an integral form, expressed in terms of $Y_X$:
\be
Y_{\Delta B}(x)
\approx -\frac{\epsilon}{2}\int_0^x dx'\,\frac{dY_X(x')}{dx'}\,\exp\left[-\int_{x'}^x\frac{dx''}{x''}\,\frac{s(x'')}{2Y_\gamma\,H(x'')}\,\langle\sigma_{\rm washout}v\rangle \prod_i Y^{\rm eq}_i(x'')\right].\label{eq:asymmetryanal}
\ee
The integrand in the exponent in eq.\ref{eq:asymmetryanal} is the washout rate $\Gamma_{\rm washout}(x)$ normalized to the Hubble scale $H(x)$,
\be\label{eq:washoutrate}
\frac{\Gamma_{\rm washout}(x)}{H(x)} = \frac{s(x)}{2Y_\gamma\,H(x)}\,\langle\sigma_{\rm washout}v\rangle \prod_i Y^{\rm eq}_i(x).
\ee
Eq.\ref{eq:washoutrate} is in analogy to the corresponding rate of WIMP annihilation, defined in:
 \be\label{eq:gammaWIMP}
\frac{\Gamma_{\rm WIMP}(x)}{H(x)}=\frac{2s(x)}{H(x)}\,\langle\sigma_{\rm ann}\,v\rangle\,Y_X(x).
\ee

Eq.\ref{eq:asymmetryanal} clearly shows that the baryon asymmetry solution can be factorized into the source term ($\frac{dY_X(x')}{dx'}$) and the exponential damping term. 

As we discussed earlier, a critical point, $x_{\rm washout} = m_X/T_{\rm washout}$, is when the washout freeze-out occurs, defined by $\Gamma_{\rm washout}/H\sim1$. Since $\Gamma_{\rm washout}(x)$ is a rapidly decreasing function at $x>1$ due to the exponential Boltzmann factor in the number density, we can model the exponential in eq.\ref{eq:asymmetryanal} with a step function, and obtain:
\be\label{eq:DLapprox}
Y_{\Delta B}(\infty) \approx -\frac{\epsilon}{2}\int_{x_{\rm washout}}^\infty dx'\,\frac{dY_X(x')}{dx} = \frac{\epsilon}{2}\left[Y_X(x_{\rm washout})-Y_X(\infty)\right],
\ee
Eq. (\ref{eq:DLapprox}) has a very clear physical interpretation: after washout scatterings freeze out, all subsequent WIMP annihilations generate a baryon asymmetry with efficiency $\epsilon$, and thus the final asymmetry $Y_{\Delta B}$ is proportional to $\epsilon$ times the total number of WIMP DM depleted though annihilations that occur after $x_{\rm washout}$ ($Y_X(x_{\rm washout})-Y_X(\infty)$).

Based on these general formulations, we can make a simple estimate of the baryon asymmetry that can be produced in this new mechanism. An upper limit can be obtained by working in the preferred weak washout regime where the mass of the exotic baryon $\psi$ satisfies $m_\psi\gtrsim m_X$. The kinematic condition for DM annihilation, $m_\psi < 2m_X$ bounds how early $x_{\rm washout}$ can be relative to $x_{\rm ann}$. For a TeV mass WIMP DM, freeze out time $x_{\rm ann}\sim 30$. Under the general assumption that both $X$ and $\psi$ have weak scale masses and interactions, $x_{\rm ann}/x_{\rm washout}\sim m_\psi/m_X$. Therefore $x_{\rm washout}\approx x_{\rm ann}(m_X/m_\psi)\gtrsim15$. Meanwhile, in this weak washout regime, washout freezes out when WIMP annihilation is still efficient, therefore $Y_X(x_{\rm washout}) \gg Y_X(\infty)$. We then obtain from (\ref{eq:DLapprox}):
\be\label{eq:asymmetryestimate}
Y_{\Delta B}(\infty) \approx \frac{\epsilon}{2}\,Y_X(x_{\rm washout})< \frac{\epsilon}{2}\,Y_X^{\rm eq}(15) \approx \epsilon\times10^{-8}.
\ee
We see that the baryon asymmetry is independent of the absolute mass $m_X$, and only depends on the ratio $x_{\rm washout} \approx m_\psi/m_X$. 
Consequently, we find the ratio of baryon to DM relic abundances:
\be
\frac{\Omega_B}{\Omega_X} = \frac{m_{\rm proton}\,Y_{\Delta B}(\infty)}{m_X\,Y_X(\infty)}\approx\frac{\epsilon}{2}\,\frac{Y_X(x_{\rm washout})}{Y_X(x_{\rm ann})}\left(\frac{\rm GeV}{m_X}\right)\lesssim\frac{\epsilon}{2}\,\frac{Y_X^{\rm eq}(15)}{Y_X^{\rm eq}(30)}\left(\frac{\rm GeV}{m_X}\right)\approx\,\left(\frac{\epsilon}{10^{-2}}\right)\left(\frac{\mathrm{TeV}}{m_X}\right)\label{eq:bdratio1}.
\ee
Therefore, with weak scale masses and O(1) couplings, the observed baryon abundance ($Y_{\Delta B}\sim10^{-10}$, $\frac{\Omega_B}{\Omega_X}\sim0.2$) is achievable in this new framework. A lower limit of the generated baryon asymmetry can also be estimated, corresponding to the disfavored strong washout case where the washout processes freeze out after WIMP annihilation freeze out, typically when $\psi$ is much lighter than $X$. The details of this estimate can be found in \cite{Cui:2011ab}, and we simply quote the result here: 
\be
\frac{\Omega_B}{\Omega_X}\gtrsim10^{-3}\times\epsilon\left(\frac{m_\psi}{m_X}\right)\left(\frac{\mathrm{TeV}}{m_X}\right).\label{eq:bdratio2}
\ee
We see that in the strong washout limit, the generated asymmetry is typically not enough to explain the observed baryon abundance, unless $m_X$ is much below weak scale.\\

\begin{figure}
\begin{center}
\includegraphics[width=8cm]{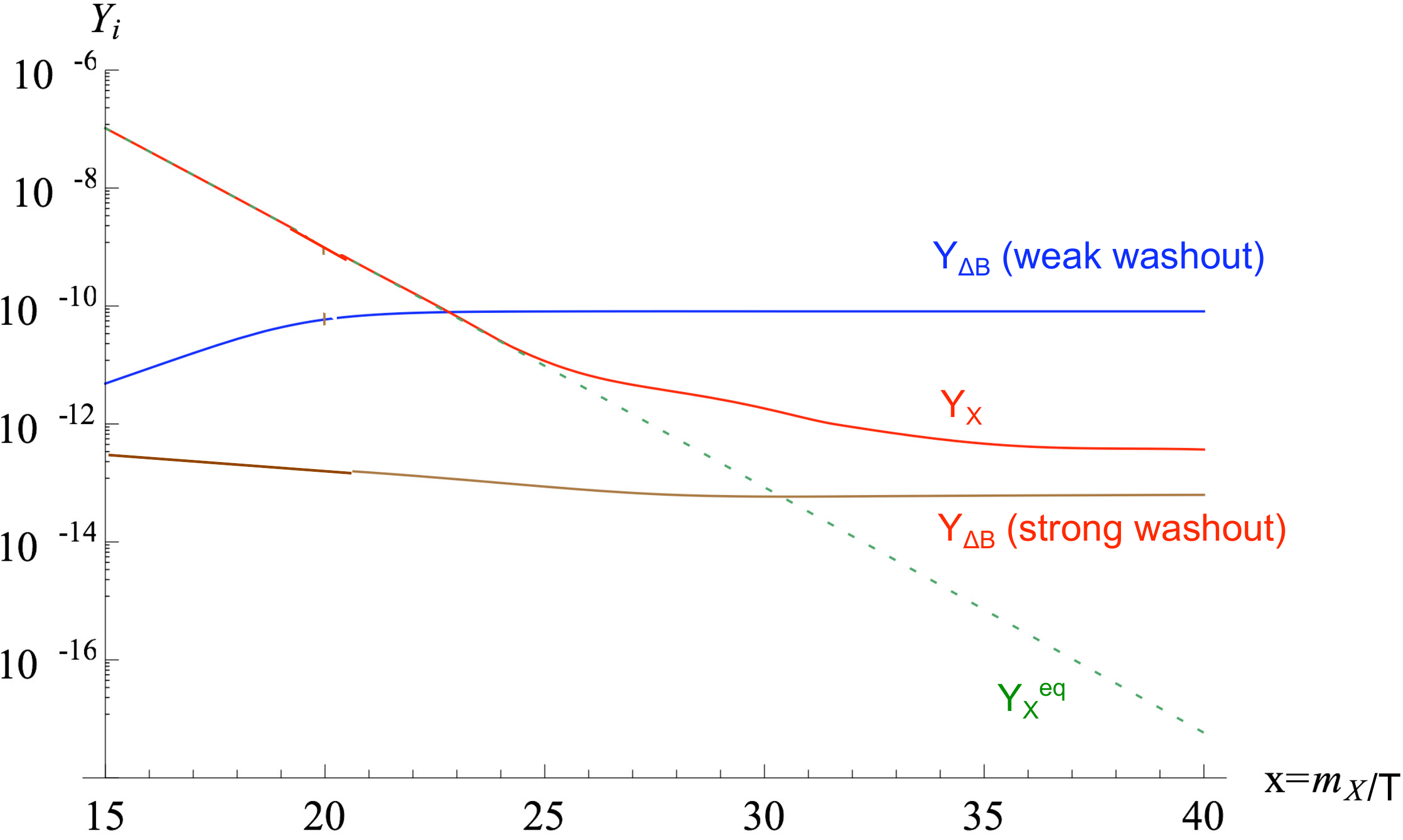}
\caption{The evolution of the number density per comoving volume for field $i$ ($Y_i$) as a function of $x=m_X/T$. The numerical solutions shown here are based on the WIMPy leptogenesis model discussed in Section \ref{sec:wimplep}, where the dominant annihilation process is $XX\rightarrow L\psi$ and the dominant washout is $L\psi\rightarrow L^\dag\psi^\dag$. The input parameters are $y_X=2.7,~ \lambda_L=0.8,~ \epsilon=0.2,~ m_X=3\rm~TeV$, and $m_S=5\rm ~TeV$. $m_\psi=4$ TeV gives the behavior when washout freezes out well before WIMP annihilation freezes out (``weak washout'').  $m_\psi=2\rm ~TeV$ gives the behavior when washout becomes ineffective subsequent to WIMP freeze-out (``strong washout'').
}\label{fig:Yplots}
\end{center}
\end{figure}

Before moving on to model examples and the related studies, in Fig.\ref{fig:Yplots} we illustrate the evolution of the abundance of DM $X$ as well as that of baryon asymmetry in the two limiting regimes of weak and strong washout. The plots are obtained by numerically solving the (exact) coupled Boltzmann equations derived for the WIMPy leptogenesis model we will demonstrate soon. The illustration based on that particular example confirms the expected features derived from our general analysis.

\subsection{Example Models}
The general mechanism of WIMPy baryogenesis can be implemented in a variety of models. Here we briefly review the essential aspects of the two example models presented in the original paper \cite{Cui:2011ab}. The first example model we will review is a leptogenesis model where WIMP directly annihilate into leptons, which require heavy WIMP such that the WIMP freeze-out occurs early enough and sufficient lepton asymmetry can be converted to baryon asymmetry before the sphaleron process shuts off around $T\sim100$ GeV. In the second model we will briefly review, WIMP directly annihilates into quarks, thus directly produce baryon asymmetry, which allows for a lighter WIMP DM.

\subsubsection{WIMP Annihilation to Leptons (WIMPy Leptogenesis) }\label{sec:wimplep}
We first consider a minimal model of WIMPy leptogenesis, where DM is a pair of Dirac fermions $X$ and $\bar X$ that annihilate to the Standard Model lepton doublet $L_i$ and exotic lepton $\psi_i$. The directly generated lepton asymmetry is converted to baryon asymmetry through an active sphaleron process. $X$ annihilates via s-wave by exchanging SM gauge singlet pseudoscalars $S_\alpha$. More than one flavour of $S_\alpha$ is necessary in order to have a physical CP violation phase in the annihilation amplitude. By gauge invariance, $\psi_i$ has charge $(2,1/2)$ under the SM EW gauge symmetry $\mathrm{SU}(2)_{\rm L}\times\mathrm{U}(1)_Y$. All the new fields are assumed to have weak scale masses. The Lagrangian is as follows:
\be\label{eq:lagrangian}
\mathcal L = \mathcal L_{\rm kin} + \mathcal L_{\rm mass} -\frac{i}{2}\left(\lambda_{X\alpha}X^2+\lambda_{X\alpha}'\bar X^2\right)S_\alpha + i\,\lambda_{L\,\alpha i}\,S_{\alpha} L_i\psi_i +\mathrm{h.c.}
\ee
where $\lambda_{L\,\alpha i}$ must be complex in order to satisfy the Sakharov CP-violation condition.

As discussed in the introductory section, in order not to over-close the universe, $\psi_i$ needs to decay to light sterile neutrino $n_i$ (which can be light enough and does not over-close the universe). Such decay can proceed through the Higgs boson portal:
\be
\Delta\mathcal L = \lambda_i\,H^\dagger n_i\psi_i+\mathrm{h.c.}
\ee
The above decay alone does not ensure that the $\psi$ asymmetry is sequestered in the sterile neutrino sector (so that there can be a net SM lepton/baryon asymmetry). Charged under the SM EW symmetry group, $\psi$ may also decay into SM anti-leptons which would cancel the SM lepton asymmetry. Additional symmetry is required in order to forbid operators that can lead to such decay. We consider a simple possibility of a $Z_4$ symmetry, with the detailed charge assignment listed in \cite{Cui:2011ab}. The SM fields are singlets under this $Z_4$, while all other new particles carry non-trivial $Z_4$ charges.

\begin{figure}[t]
\begin{center}
\includegraphics[height=2.2cm]{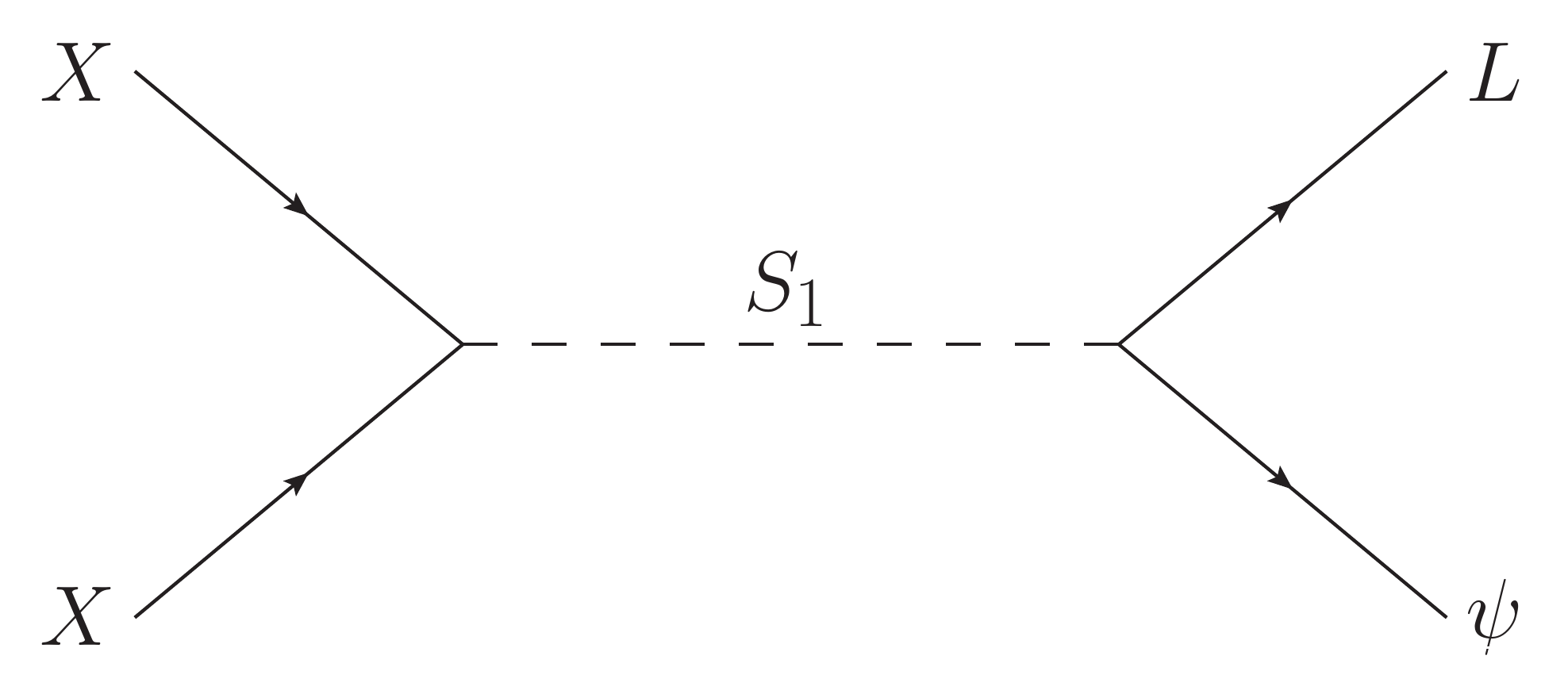}\hspace{0.7cm}\\
\includegraphics[height=2.2cm]{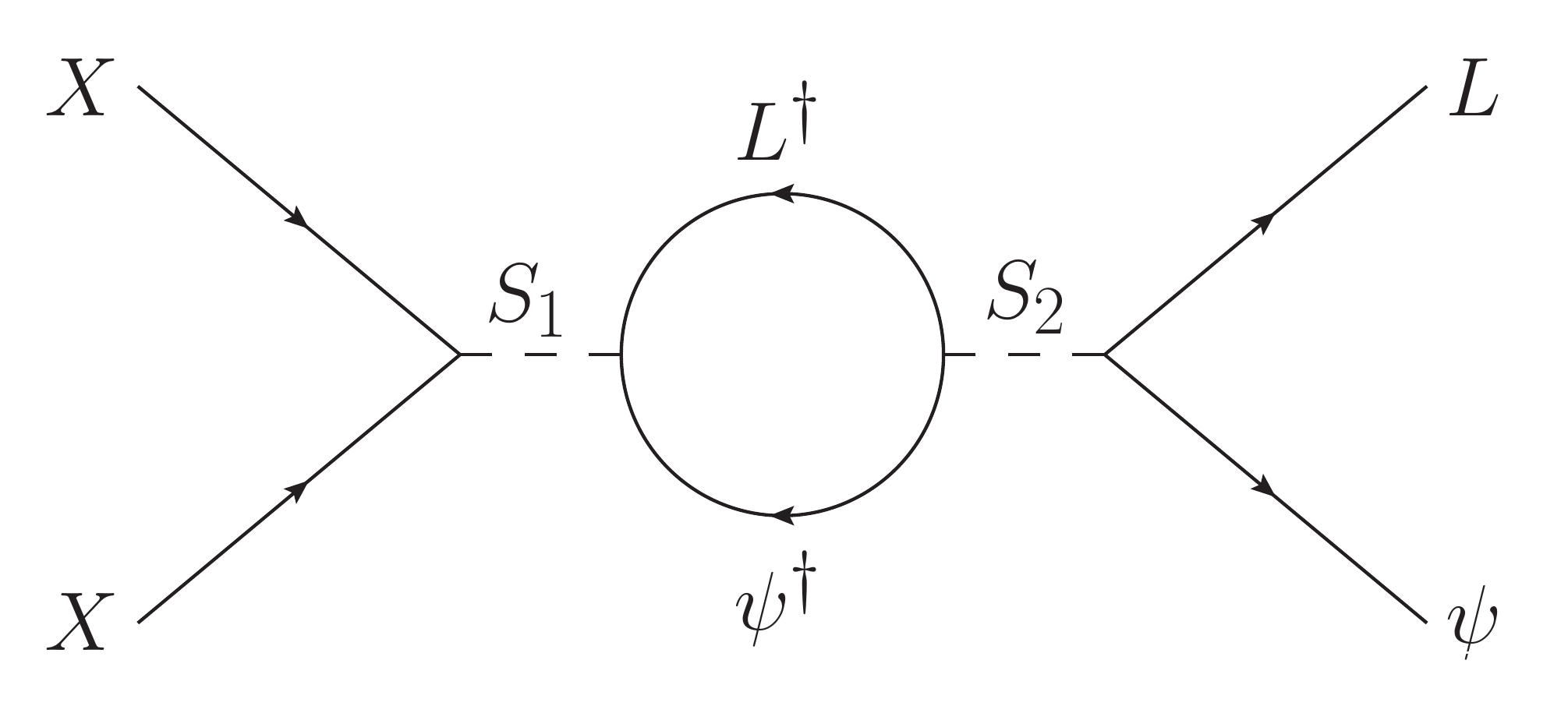}\hspace{0.7cm}
\includegraphics[height=2.2cm]{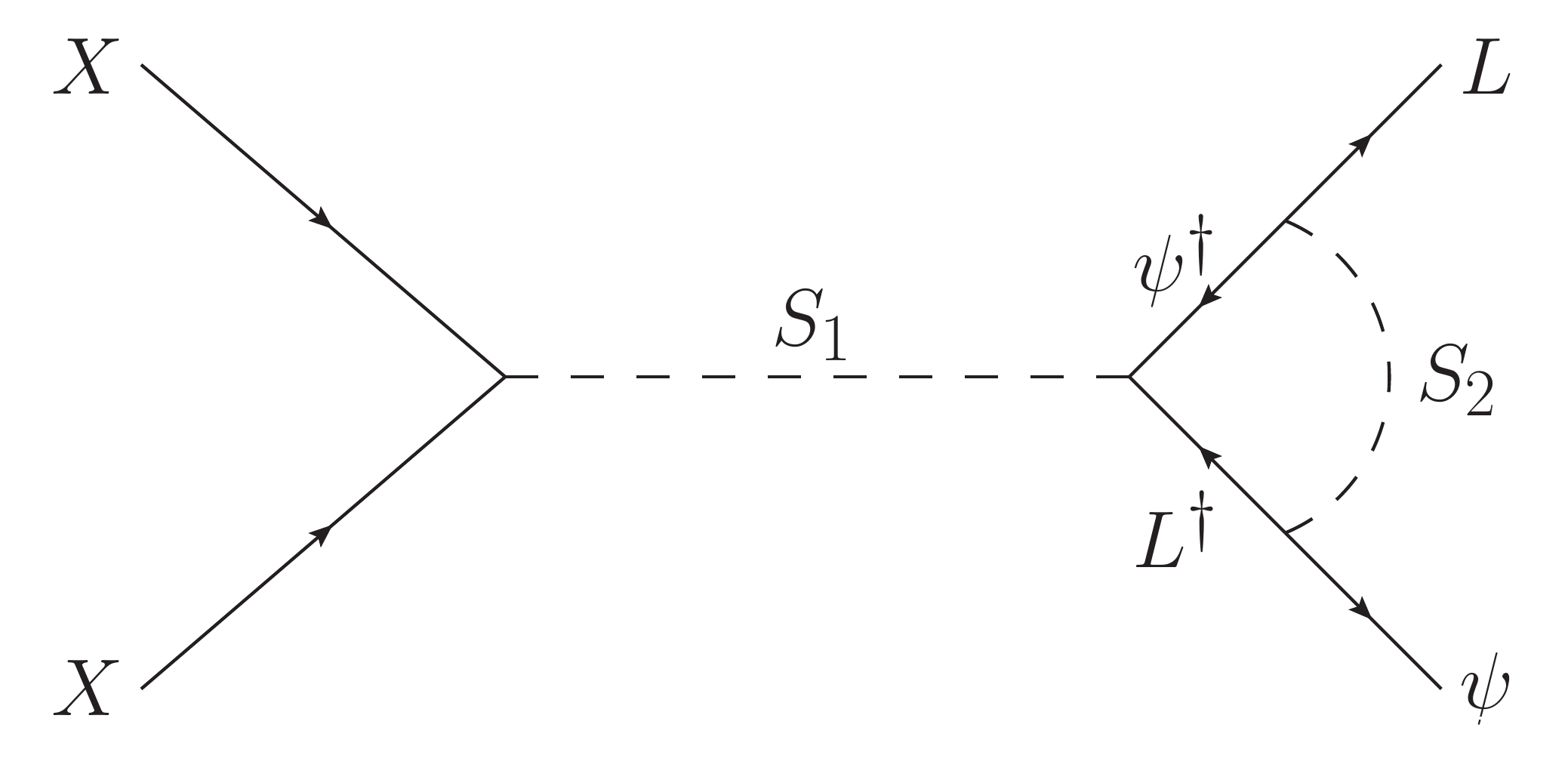}
\caption{Diagrams of tree and loop contributions to the $XX$ annihilation cross section. The difference between these rates and their conjugates generates a lepton asymmetry.}
\label{fig:asymmetry}
\end{center}
\end{figure}

The CP asymmetry $\epsilon$ of the L-violating WIMP annihilation is defined as follows:
\be
\epsilon= \frac{\sigma(XX\rightarrow \psi_i L_i)+\sigma(\bar X\bar X\rightarrow \psi_i L_i)-\sigma(XX\rightarrow\psi_i^\dagger L_i^\dagger)-\sigma(\bar X\bar X\rightarrow\psi_i^\dagger L_i^\dagger)}{\sigma(XX\rightarrow \psi_i L_i)+\sigma(\bar X\bar X\rightarrow \psi_i L_i)+\sigma(XX\rightarrow\psi_i^\dagger L_i^\dagger)+\sigma(\bar X\bar X\rightarrow\psi_i^\dagger L_i^\dagger)}.
\ee
A nonzero $\epsilon$ arises from the interference between the tree-level process and the corresponding loop-level processes, which are shown in Fig.\ref{fig:asymmetry}. One can compute the asymmetry in this model and the result is as follows:
\be\label{eq:epsilonexpr}
\epsilon\approx-\frac{1}{6\pi}\,\frac{\mathrm{Im}(\lambda_{L1}^2\lambda_{L2}^{*2})}{|\lambda_{L1}|^2}\,\frac{(2m_X)^2}{m_{S2}^2}\left[7-15\left(\frac{m_\psi}{2m_X}\right)^2+9\left(\frac{m_\psi}{2m_X}\right)^4-\left(\frac{m_\psi}{2m_X}\right)^6\right].
\ee

%-------------------------------------------------------------------
%-------------------------------------------------------------------
As discussed in more details in the original paper \cite{Cui:2011ab}, in this model a complementary source of baryon asymmetry can come from $S_\alpha\rightarrow \psi_iL_i$ decay, which is in direct analogy to conventional leptogenesis from out-of-equilibrium decay. The annihilation is the dominant source for baryogenesis when $m_X<m_S$.\\
      As emphasized in the general analysis section, in WIMPy barogenesis, the amount of baryon asymmetry is sensitive to washout processes and when they become inefficient (freeze-out). We illustrate the leading washout scatterings in Fig.  At $T\ll m_\psi$, $3\rightarrow3$ washout process $LHn^\dag\rightarrow L^{\dag}H^*n$ can dominate over these $2\rightarrow2$ processes.
\begin{figure}[t]
\begin{center}
\includegraphics[width=2.5cm]{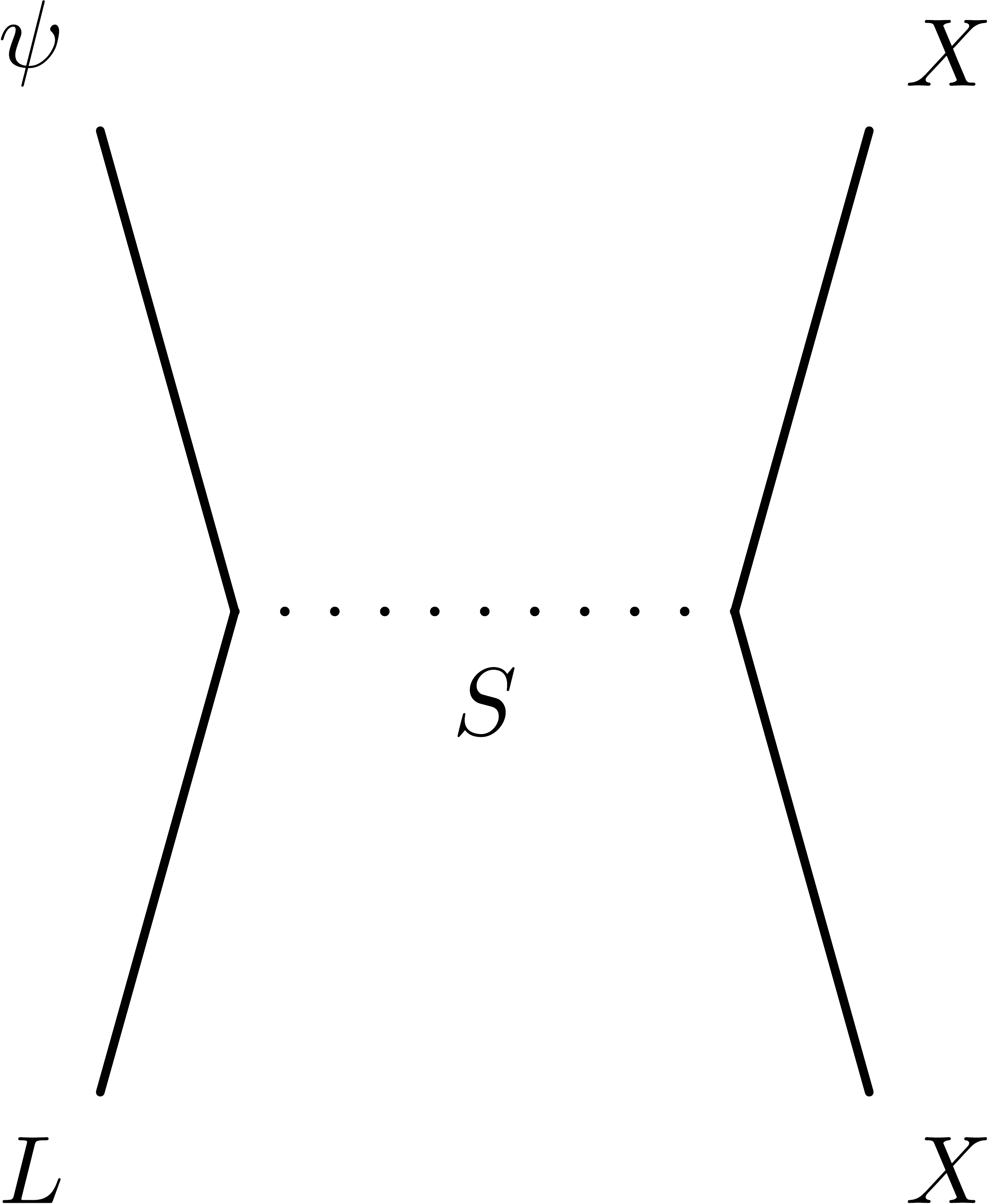}\hspace{1cm}
\includegraphics[width=2.5cm]{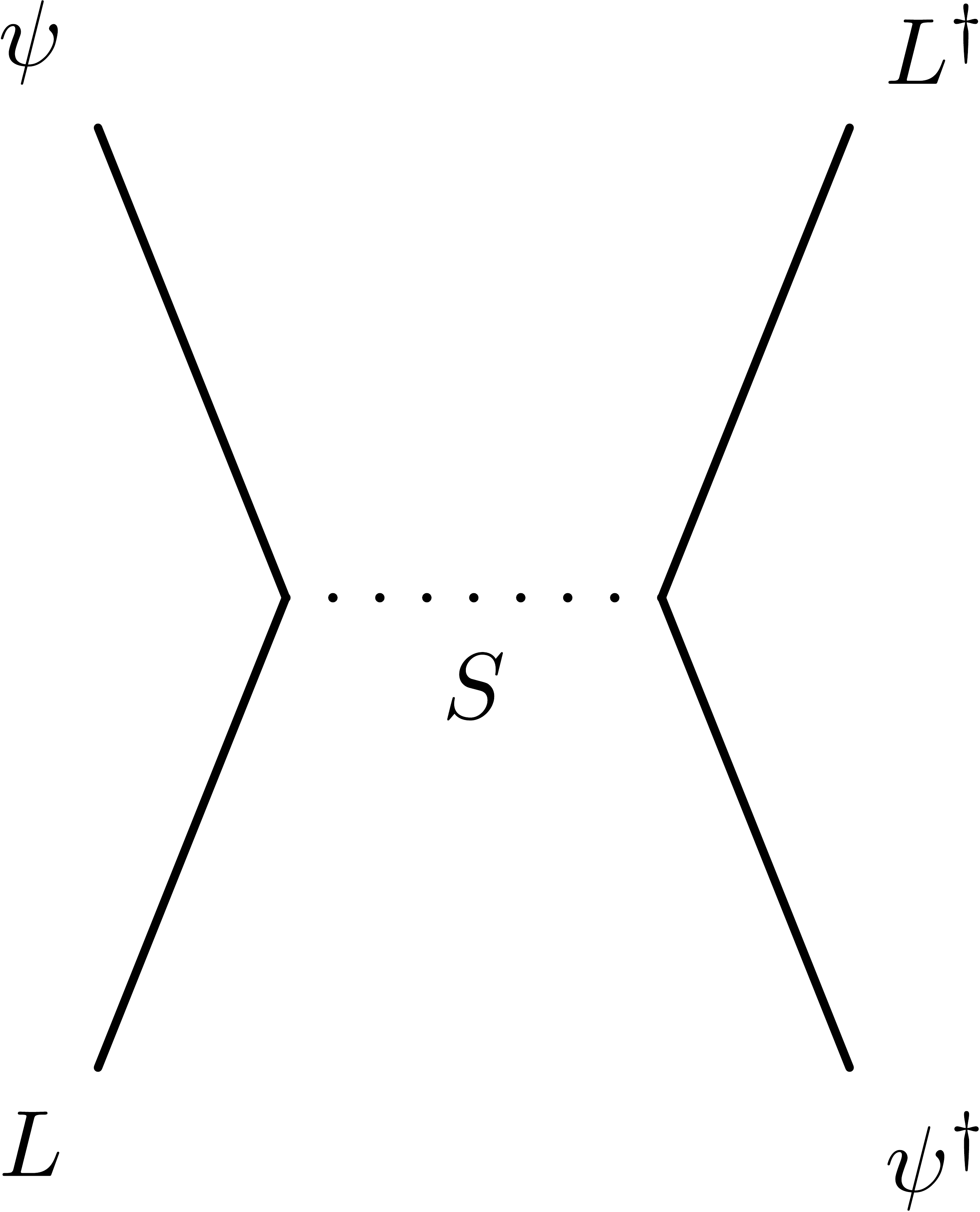}
\\
\vspace{0.5cm}
\includegraphics[width=2.5cm]{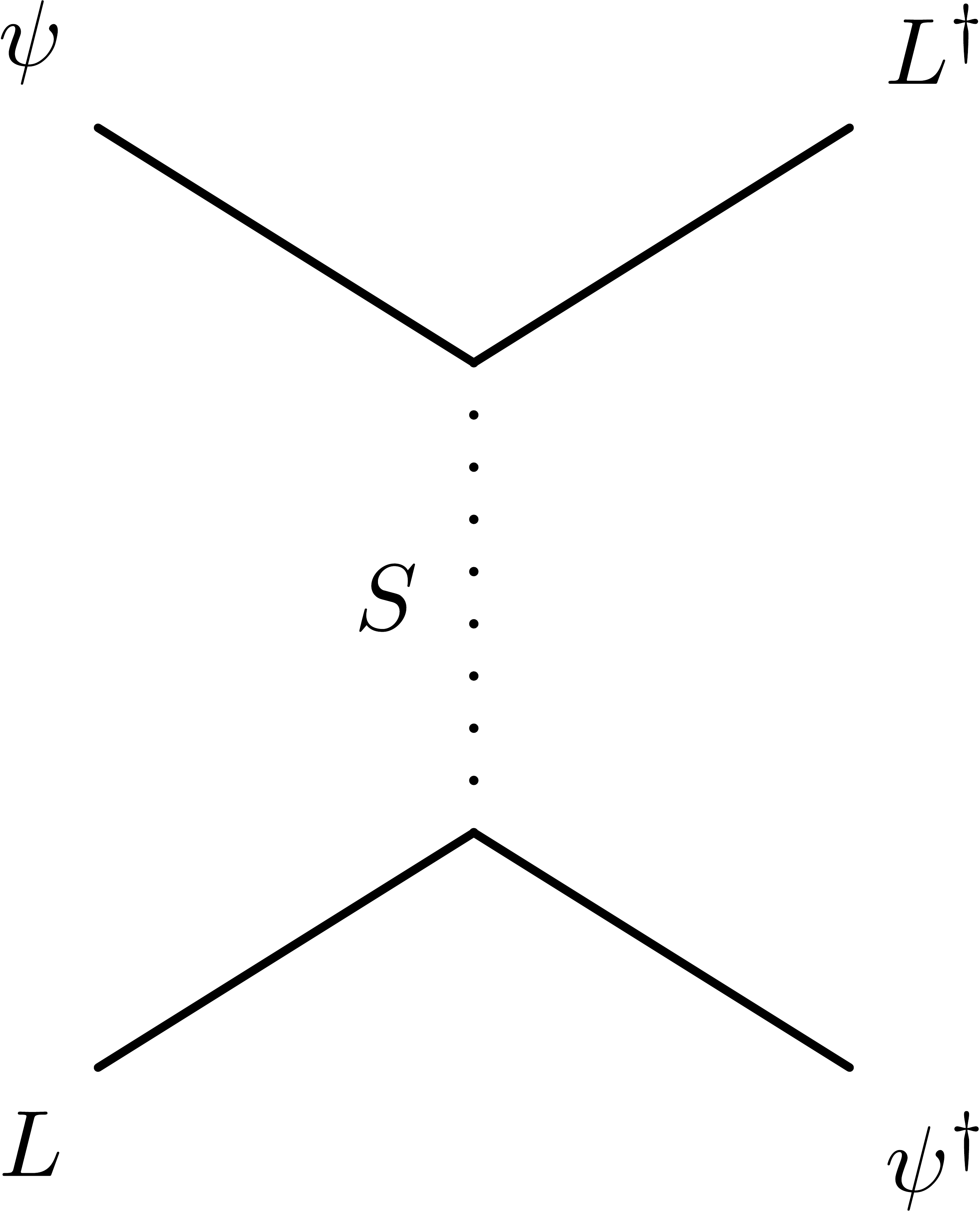}\hspace{1cm}
\includegraphics[width=2.5cm]{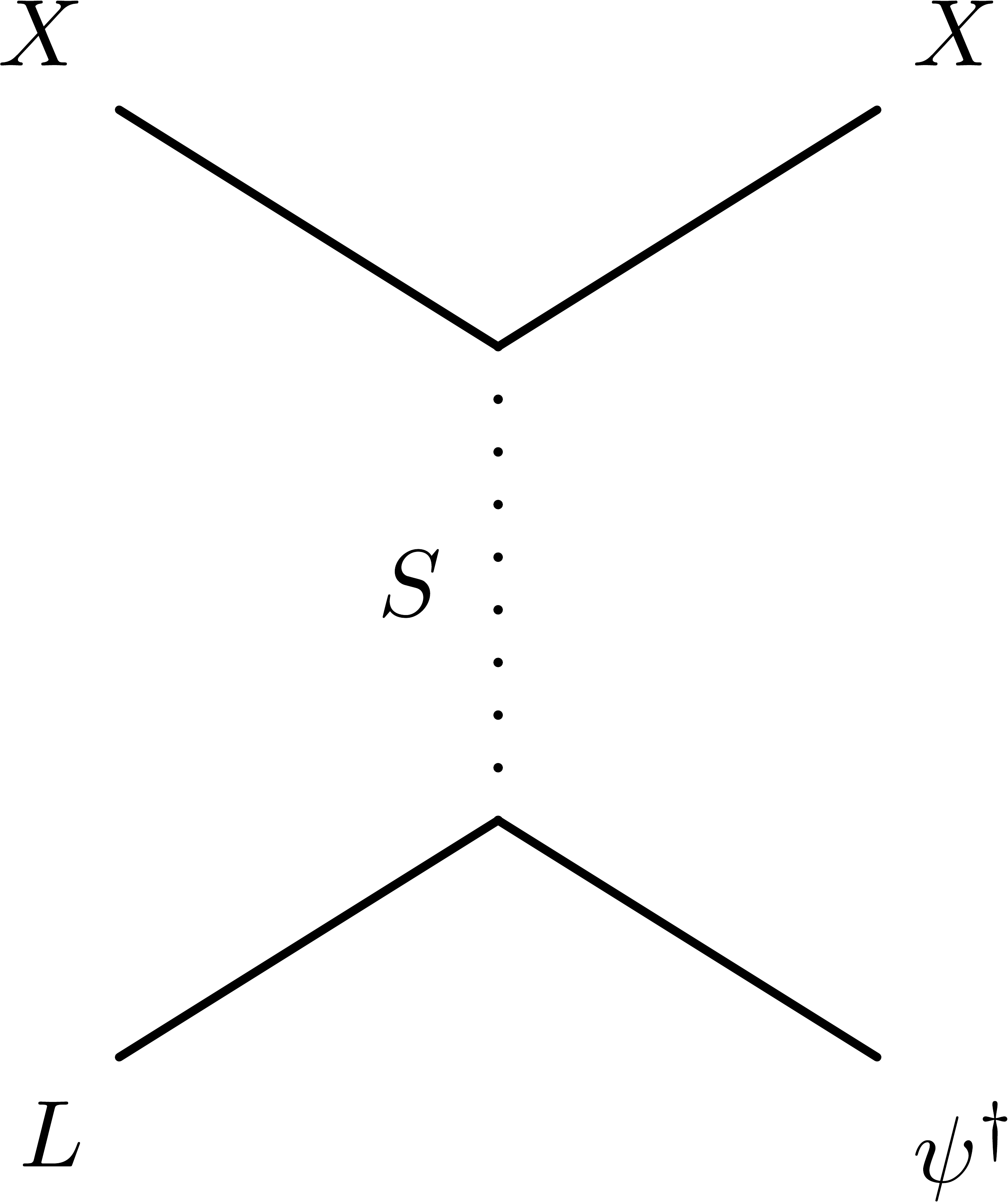}\hspace{1cm}
\includegraphics[width=2.5cm]{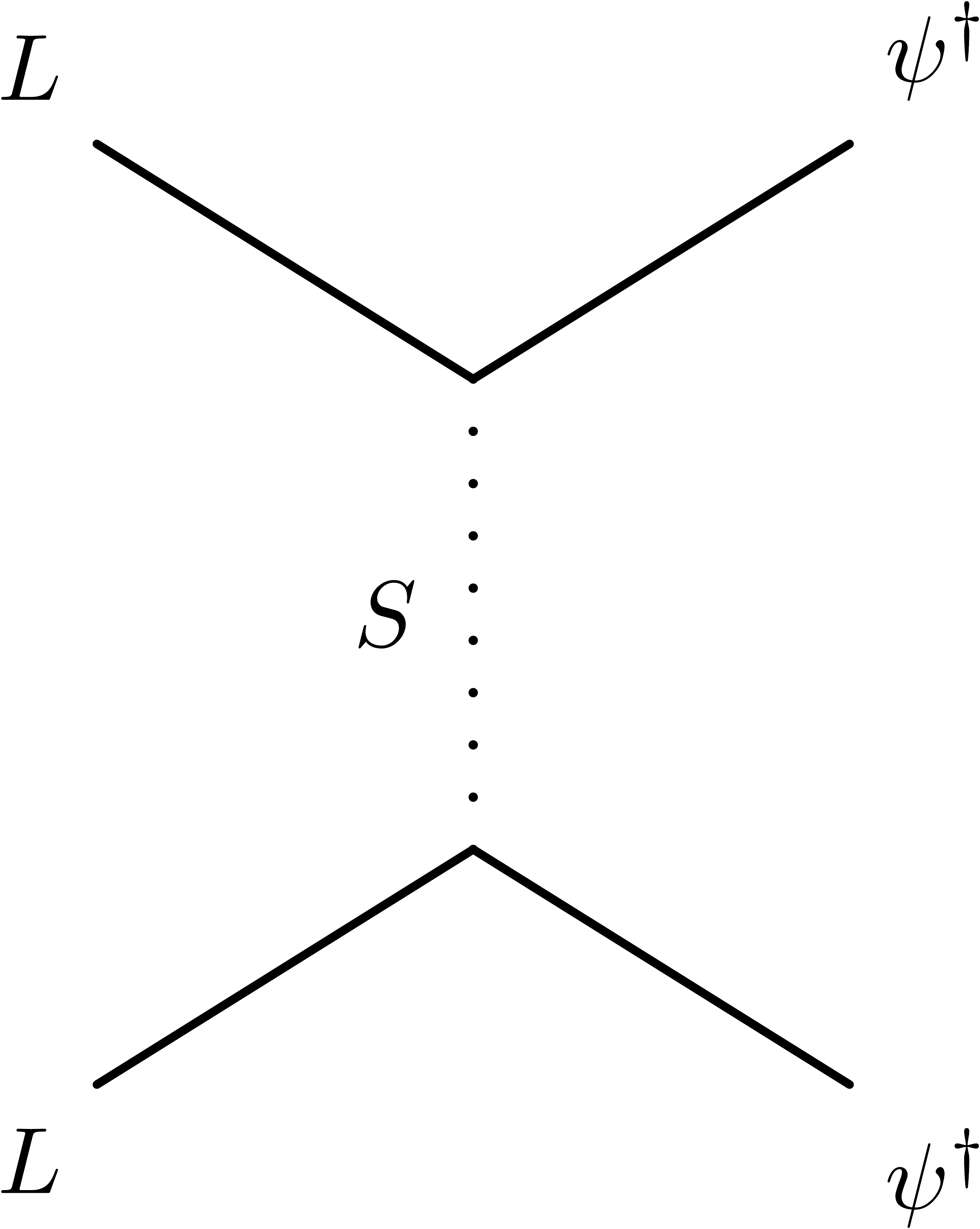}
\caption{Diagrams leading to washout of the lepton number from (top row) $s$-channel and (bottom row) $t$-channel scatterings.}
\label{fig:washout}
\end{center}
\end{figure}

\begin{figure}[t]
\begin{center}
\includegraphics[width=7cm]{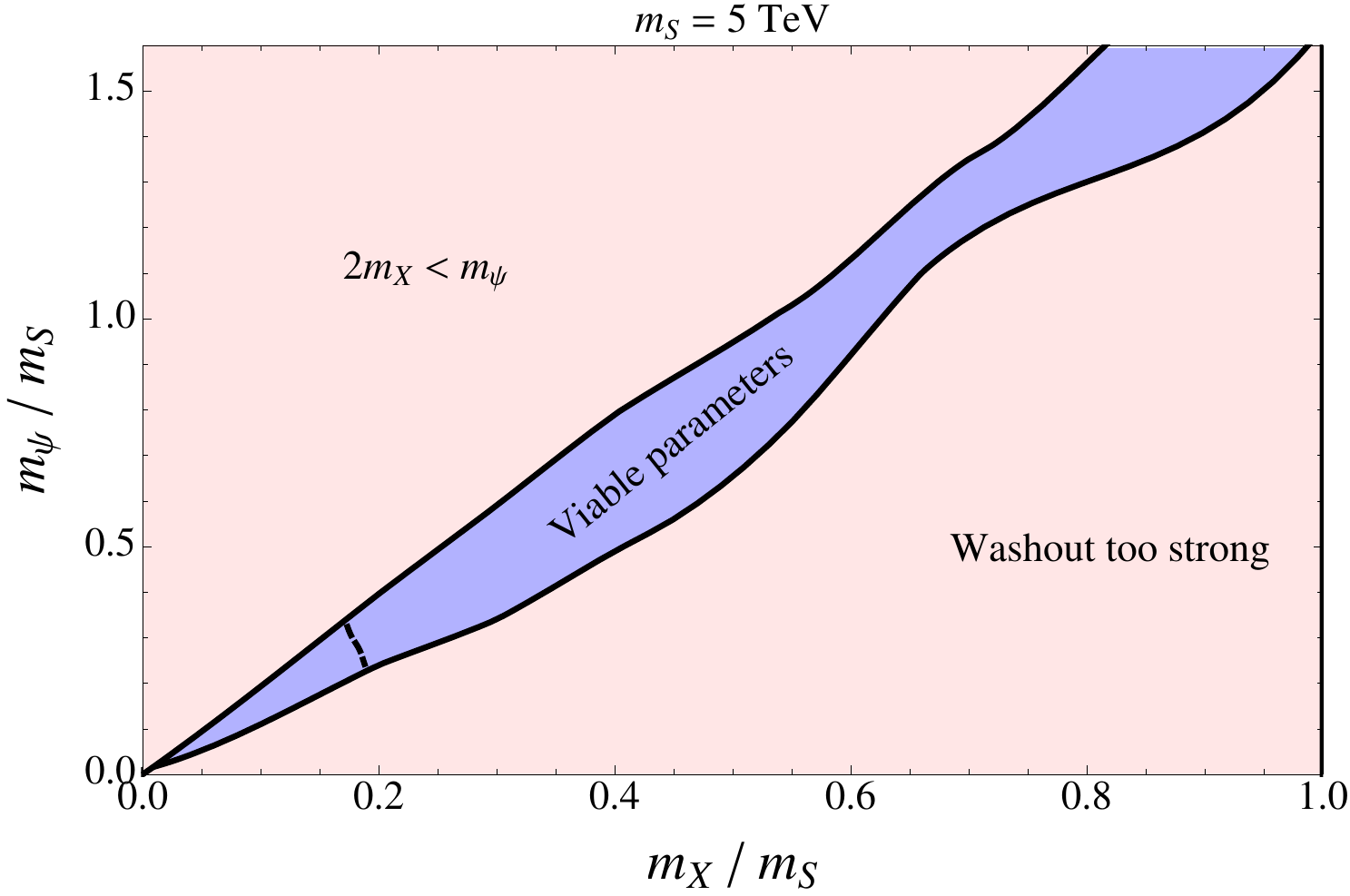}
\caption{Regions in the $m_X$-$m_\psi$ plane with the correct WIMP relic density and baryon asymmetry from WIMPy leptogenesis, with $m_S=5$ TeV and some choice of perturbative couplings. The masses giving both observed abundances are shown in blue (middle stripe).  We plot the ratios $m_X/m_S$, $m_\psi/m_S$  to show the relationship between the $X$ and $\psi$ masses and the  mediator scale $m_S$.  The excluded regions are shown in red: the upper region is not viable because $2m_X<m_\psi$ and the thermal annihilation cross section is  Boltzmann-suppressed, while the lower region has $Y_\psi$ too large to prevent rapid washout of the asymmetry. The dashed line indicates the lower boundary of allowed $m_X$ and $m_\psi$; below the line, the electroweak phase transition occurs before the baryon asymmetry is large enough to account for the observed value.  For $m_X/m_S>1$, the asymmetry is dominated by $S$ decay. }
\label{fig:mxmpsimain}
\end{center}
\end{figure}

Taking into account all these relevant processes, we derive the exact Boltzmann equations for this model in \cite{Cui:2011ab}, and obtain the numerical solutions for the DM and baryon relic abundances. Here we show the results with the benchmark point in Fig.\ref{fig:mxmpsimain}. As expected from our general analysis, the mass parameter is restricted to a relatively narrow region where $m_\psi/2\lesssim m_X\lesssim m_\psi$, due to the washout effect.

%%%%%
\subsubsection{WIMP Annihilation to Quarks}
In this type of model, WIMP DM annihilate to baryons directly, therefore the cosmological lower bound of $m_X\gtrsim$ TeV based on sphaleron consideration does not apply here, unlike in the case of WIMP leptogenesis. Nonetheless, these models involve new particles charged under SM color group, which can be copiously produced at colliders such as the LHC, and thus are subject to strong constraints unless they are massive enough. The model content is similar to the leptogenesis model that we just discussed: vectorlike gauge singlet dark matter $X$ and $\bar X$, singlet pseudoscalars $S_\alpha$, and vectorlike exotic quark color triplets $\psi_i$ and $\bar\psi_i$. The Lagrangian is
\be\label{eq:lagrangianquark}
\mathcal L = \mathcal L_{\rm kin} + \mathcal L_{\rm mass} -\frac{i}{2}\left(\lambda_{X\alpha}X^2+\lambda_{X\alpha}'\bar X^2\right)S_\alpha + i \lambda_{B\,\alpha}\,S_\alpha\bar u\psi.
\ee
The exotic quark $\psi$ can not be stable in order not to overclose the universe, and an additional symmetry such as $Z_4$ is needed to prevent direct decay of  $\psi$ to SM anti-baryons through a $QH\bar\psi$ term and thus erase the produced asymmetry. We consider two possible patterns of $\psi$ decay:
\begin{enumerate}
\item Decay through B-conserving interaction: $\psi_i$ decays to light, B-number-carrying SM gauge singlets $n_i$ and a SM antiquark. The additional terms in the Lagrangian are:
\be\label{eq:lagrangianquark1}
\Delta\mathcal L = \lambda_i\,\bar\psi_i\,\bar d_i\,\phi^*+\lambda_i'\,\phi\,\bar d_i\,n_i+\mathrm{h.c.}
\ee
where scalar $\phi$ has the SM $SU(3)_C\times SU(2)_L\times U(1)_Y$ charges $(3,1,-1/3)$. Apparently, such scalar can be naturally realized as a right-handed squark in supersymmetric models.
\item Decay through B-violating interaction: $\psi_i$ decays to two SM antiquarks and a Majorana singlet $n$ which does not carry B-number. Again the Lagrangian would include a color triplet scalar which can be identified as squark in supersymmetry (SUSY) models. The additional terms in the Lagrangian are:
\be\label{eq:lagrangianquark2}
\Delta\mathcal L =\lambda\,\epsilon^{ijk}\,\bar\psi_i\,\bar d_j\,\tilde d^*_k+\lambda_i'\,\bar d_i\,\tilde d_i\,n+\mathrm{h.c.}
\ee
\end{enumerate}
Apparently in this scenario, the singlet $n$ can be realized in SUSY models as neutralino, while the first term in eq.\ref{eq:lagrangianquark2} resembles the $UDD$ type of R-parity violating interaction in SUSY framework. The detailed analysis of this model can be found in \cite{Cui:2011ab}, with qualitative results similar to that of the leptogenesis model. \\

There have been further efforts in building concrete models that realize the general idea proposed in \cite{Cui:2011ab}. These recent studies can be found in \cite{Bernal:2012gv, Bernal:2013bga, Kumar:2013uca, Stengel:2014jta, Racker:2014uga}.

\subsection{Phenomenology}
With new particles at weak scale, WIMPy baryogenesis models have interesting phenomenological implications for a variety of experiments. The detailed constraints and potential signals for the above two simple models have been discussed in length in \cite{Cui:2011ab}, and further studies can be found in \cite{Bernal:2012gv}. In this review letter, we just summarize the main results. At dark matter direct detection experiments, certain parameter region of the direct baryogenesis model can be probed in the near future, while the signal from the leptogenesis model is hard to observe due to the 1-loop suppressed scattering off electrons. The direct baryogenesis model can give rise to potentially interesting induced nucleon decay, similar to the Hylogenesis models \cite{Davoudiasl:2010am}, but the signal is much suppressed with weak scale masses. At colliders, the exotic doublet leptons in WMPy leptogenesis can be pair-produced through electroweak interaction, with the cascade decay as: $\psi^0\psi^0\rightarrow h h + \cancel E_{\rm T}\rightarrow 4b(4j)+ \cancel E_{\rm T}$  $\psi^+\psi^-\rightarrow W^+ W^-+\cancel E_{\rm T}$. The exotic quark in the direct baryogenesis models can be produced through strong interaction with the cascade decay: $pp\rightarrow\psi\bar\psi\rightarrow 4j+ \cancel{E}_{\mathrm T}$. Provided that the masses are not too heavy, these exotic leptons or baryons may be within the reach of the 14 TeV LHC, in particular with targeted searches. For these minimal models, the resultant electron or neutrino electric dipole moment (EDM) only arises at 3-loop and thus much suppressed, as the CP-violation only arises in the coupling to one chirality of SM fermion, similar to the situation with the SM.  Note that some of the non-observable signals can be observable in variational or more elaborate models that realize the WIMPy baryogenesis mechanism. \\

\textbf{Summary:} WIMPy baryogenesis provides a new mechanism to simultaneously produce WIMP DM and baryon asymmetry from WIMP annihilation around the freeze-out time, and leads to a variety of interesting phenomenology. The slight drawback of this scenario is that the generated baryon asymmetry is sensitive to model details, in particular the washout processes, and a peculiar mass relation is required in order to suppress the washout. The connection to the WIMP miracle is thus less robust at a quantitative level. This motivated the proposal of an alternative WIMP-triggered baryogenesis mechanism that we will discuss next, which predicts the baryon asymmetry in a more robust way, insensitive to model details. 

\section{Baryogenesis from Metastable WIMP Decay }
\subsection{General Idea and Formulation}
We start this section by introducing a generalized concept of the WIMP-type of new particle. As discussed in the Introduction, solutions to the electroweak hierarchy problem typically involve new weak scale mass particles $\chi$ with interaction strength $G_\chi\sim G_F$, that is, ``WIMP''. Often times, ``WIMP'' is automatically associated with a stable WIMP which can be dark matter candidate. However, generally there can be an array of WIMPs, some of which are stable, some of which decay promptly, while some of which can have a long yet finite lifetime, i.e. metastable. From the particle physics model-building point of view, such diversity of lifetimes is not an ad-hoc complexity, rather it can naturally arise from symmetry protection and mass hierarchy. We have seen familiar examples from the Standard Model: the SM fermions have hierarchical masses which may result from a flavour symmetry; massive fermions such as $b$-quark and $\tau$-lepton can be metastable and leave a macroscopic-scale decay track in collider experiments, due to the much heavier $W$ boson that mediates their decay. In the well-studied Minimal Supersymmetric Standard Model (MSSM), the lightest neutralino, a classic example of a WIMP particle, is stable when R-parity is an exact symmetry, yet can decay when R-parity is violated, with a lifetime dependent on the magnitude of the R-parity violation.

Now with the above generalized concept that WIMP can have a diverse lifetime, let us recall the thermal history of a WIMP particle as we discussed in Section.\ref{sec:warmup}. A WIMP particle stays in thermal equilibrium in the hot early universe, then departs from equilibrium during the freeze-out stage when its annihilation rate falls below the Hubble expansion rate. If a WIMP is stable, after the thermal-freezeout, its comoving relic abundance essentially ``freezes in'' and approaches the current day value. In this scenario, nothing nontrivial occurs in between thermal freeze out and today, and the stable WIMP remains as DM today. However, considering that it is a long cosmic time span between the WIMP freeze-out ($10^{-12}-10^{-6}$ seconds after the Big-Bang, depending on WIMP mass and annihilation cross section) and the current day time ($10^{18}$ seconds after the Big-Bang), during which many significant transitional events may occur. The familiar cosmic history of a stable WIMP particle is in fact only the most trivial/simple possibility. As we just discussed, a WIMP particle can generally be unstable. A particularly interesting case that we will consider is a metastable WIMP that first undergoes thermal freeze-out, then at a later time it decays to SM states in a pattern that violates B- (or L-) number and CP symmetry. Due to the earlier stage of thermal freeze-out, at the time of its decay, the WIMP number density has well departed from equilibrium distribution. Apparently this scenario naturally provides all the Sakharov conditions for a viable baryogenesis mechanism. Now let us discuss the relevant processes in more detail, following the time order.

{\underline{Stage-1: WIMP freezeout}\\
At this early stage the evolution of the baryon parent WIMP $\chi_B$ is just like that of a familiar WIMP DM. The freeze-out temperature $T_f$, and the comoving number density $Y_\chi(x_f)$ at the end of the freeze-out can be estimated using the same eqs.\ref{omegawimp}. Again these quantities are neatly determined by the WIMP mass and its thermal annihilation cross section. The difference from a stable WIMP is that, the non-trivial evolution of $Y_{\chi_B}$ does not end at freeze-out. $Y_\chi(x_f)\equiv Y_{\chi_B}^{\rm ini}$ sets the initial condition for baryogenesis that occurs after freeze-out, which we will discuss next. The WIMP miracle prediction for the relic abundance of WIMP DM as in eq.\ref{omegawimp} is still meaningful in this case. It should now be understood as the ``would-be'' relic abundance of the $\chi_B$ in the limit when $\chi_B$ is stable (i.e. lifetime $\tau\rightarrow\infty$), $\Omega_{\chi_B}^{\tau\rightarrow\infty}$.      \\

{\underline{Stage-2: Baryogenesis}}\\
Now consider $\chi_B$ to undergo B- and CP-violating decay after its thermal freeze out but before Big Bang Nucleosynthesis (BBN), i.e. $1~\rm MeV\sim T_{\rm BBN}<T_{\rm dec}<T_f$. With this general assumption, we can treat freezeout and baryogenesis as decoupled processes, which simplifies the analysis and leads to a robust result insensitive to model details, and at the mean time retain the success of standard BBN theory. Baryogenesis through massive particle decay is a well-studied subject, with the typical Boltzmann equations given in, e.g. \cite{Kolb:1979qa}. Assuming each decay violates B-number by 1 unit, we solve the Boltzman eq. for our case, and obtain the co-moving baryon asymmetry density today as:
\bea\nonumber
Y_{{B}}(0)&=&\epsilon_{\rm CP}\int_0^{T_{\rm D}}\frac{dY_{\chi_B}}{dT}\exp\left(-\int^{T}_0 \frac{\Gamma_{\rm W}(T')}{H(T')}\frac{dT'}{T'}\right)dT\\ &+& Y_B^{\rm ini}\exp\left(-\int^{T_{\rm \rm ini}}_0 \frac{\Gamma_{\rm W}(T)}{H(T)}\frac{dT}{T}\right), \label{Bsol}
\eea
where we used current-day cosmic temperature $T_0\approx 0$, $\epsilon_{\rm CP}$ is CP asymmetry in $\chi_B$ decay, $\Gamma_{\rm W}$ is the rate of $\cancel{B}$ washout processes. $Y_B^{\rm ini}$ represents possible pre-existing B-asymmetry, which we assume to be $0$.
As we will explain more along with the example models, the weak-washout condition, i.e. $\Gamma_{\rm W}<H$ at the time of WIMP decay can generally be satisfied in this mechanism since the baryon-asymmetry generating decay can occur well below weak scale temperature when major washout processes are already inefficient. With weak washout, i.e., $\Gamma_{\rm W}<H$, the exponential factor in eq.(\ref{Bsol}) can be dropped. Then combining eqs.(\ref{omegawimp},\ref{Bsol}) and our earlier definition of $\Omega_{\chi_B}^{\tau\rightarrow\infty}$, we obtain our central result which predicts current-day baryon abundance:
\be
  Y_{{B}}(0)\simeq\epsilon_{\rm CP} Y_{\chi_B}(\Tf), ~~ \Omega_B(0)=\epsilon_{\rm CP}\frac{m_p}{m_{\chi_B}}\Omega_{\chi_B}^{\tau\rightarrow\infty}\label{omegab}.
\ee
As long as baryogenesis occurs well before BBN, the symmetric component of baryons produced from the WIMP decay can be efficiently depleted by thermal annihilation among baryons and anti-baryons, and the final baryon abundance is dominated by the asymmetry as given in eq.(\ref{omegab}). From eq.\ref{omegab} we see that the prediction of $\Omega_B$ is insensitive to washout process details or the precise lifetime $\tau$, as long as the WIMP survives thermal freeze out. Note that eq.\ref{omegab} takes the form of the WIMP miracle, except for the extra factor $\epsilon_{\rm CP}\frac{m_p}{m_{\chi_B}}\sim10^{-4}-10^{-3}$ for weak scale $\chi_B$ and $O(1)$ couplings and phases. As we emphasized in the Introduction, WIMP miracle prediction for (would-be) relic abundance naturally has a variation range of a few orders of magnitude, depending on masses and couplings. Therefore assuming another species of WIMP $\chi_{\rm DM}$ which is the stable DM protected by an exact symmetry, the observed $\frac{\Omega_B}{\Omega_{\rm DM}}\approx\frac{1}{5}$ can readily arise from $O(1)$ difference in masses and couplings associated with the two WIMP species $\chi_{\rm DM}$ and $\chi_B$. This novel baryogenesis mechanism thus provides a new path addressing the cosmic coincidence of DM-baryon abundances.\\

We summarize the key processes in this new baryogenesis mechanism in Fig.\ref{fig:cartoon}.
\begin{figure}
   \begin{center}
        \includegraphics[height=60mm]{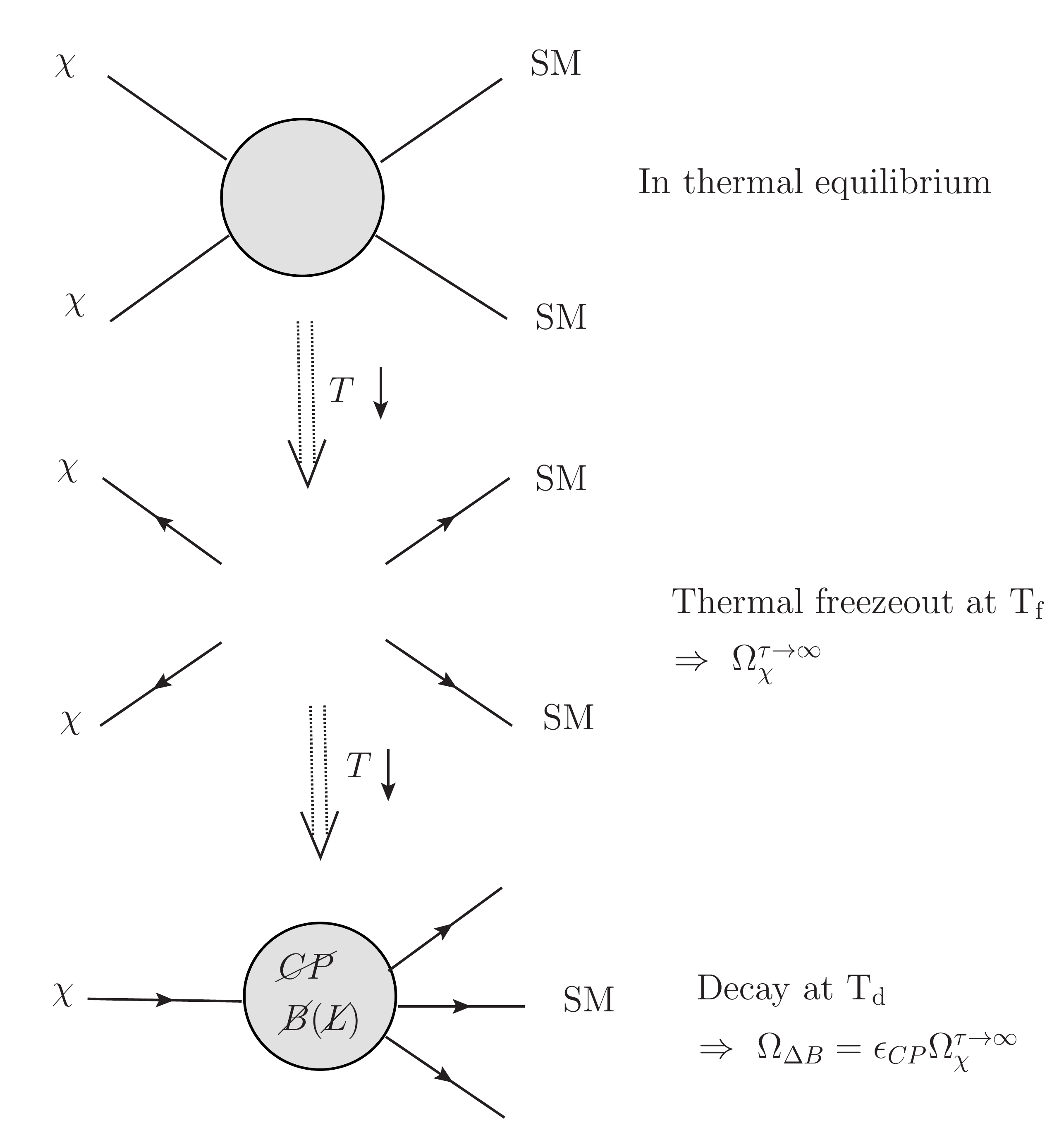} 
     \end{center} 
        \caption{Illustration of the cosmic evolution of WIMP baryon parent $\chi$. The dashed double arrow indicates the arrow of time, along which the temperature drops.}  
        \label{fig:cartoon}   
\end{figure}

\subsection{Example Models}
In this section we will review the essential aspects of the two types of models studied in the original papers \cite{Cui:2012jh, Cui:2013bta}: a minimal model based on the 2-body decay of a metastable WIMP and its implication in supersymmetric (SUSY) framework where all new particles natural lie around weak scale; then an embedding in the so-called mini-split SUSY where new scalars (except for the Higgs boson) have mass well above weak scale, in which case 3-body decay of bino realizes baryogenesis with the minimal SUSY spectrum. Following these simple examples, there have been interesting studies of variational or more extended models, some including gravitino or axino \cite{Sorbello:2013xwa, Arcadi:2013jza, Monteux:2014hua, Arcadi:2015ffa}.

\subsubsection{A Minimal Model and Its Embedding in Natural SUSY}
A minimal model can be realized by extending the SM Lagrangian with the following terms, which contain sources of baryon-number and CP violations:
 \bea
\Delta\mathcal{L}&=&\nonumber\lambda_{ij}\phi d_id_j + \varepsilon_i \chi\bar{u}_i\phi +M_{\chi}^2\chi^2 + y_i\psi\bar{u}_i\phi +M_{\psi}^2\psi^2\\ &+& \alpha\chi^2S + \beta|H|^2 S + M_S^2S^2 + \rm h.c., \label{minimalmodel}
\eea
where all couplings can be complex, $H$ is the SM Higgs boson; $d_i$ and $u_i$ are right-handed SM quarks with flavour indices $i=1,2,3$; $\phi$ is a di-quark scalar with same SM gauge charge as $u$;
$\chi$ and $\psi$ are SM singlet Majorana fermions, and $S$ is a singlet scalar. $\chi$ is the shorthand for the baryon's WIMP parent, $\equiv\chi_B$ that is defined earlier. $\varepsilon_i\ll1$ are formal small parameters leading to long-lived $\chi$. We need $10^{-13}\lesssim \varepsilon_i\lesssim10^{-8}$ in order for the decay to occur in the preferred range $1~\rm MeV\sim T_{\rm BBN}<T_{\rm dec}<T_f$. Such small couplings are technically natural, and can originate from an approximate $\chi$-parity symmetry.

In this model, baryogenesis is triggered by out-of-equilibrium decay $\chi\rightarrow\phi^* {u}$, followed by the prompt decay $\phi\rightarrow{d}{d}$. CP asymmetry $\epsilon_{\rm CP}$ in $\chi$ decay comes from the $\psi$-mediated interference between tree-level and loop diagrams as shown in Fig.(\ref{fig:decaycpv}). In the case of $M_{\psi}>M_{\chi}$, we obtain:
\be
 \epsilon_{\rm CP}\simeq\frac{1}{8\pi}\frac{1}{\sum_i|\varepsilon_i|^2}{\rm Im}\left\{\left(\sum_i\varepsilon_iy_i^{*}\right)^2\right\}\frac{M_{\chi}}{M_{\psi}}\label{cpv},
\ee
which is non-zero for generic complex couplings. Apparently in order to have a large $\epsilon_{\rm CP}$, we need  $y_i\sim O(1)$ for at least one flavor $i$. 
\begin{figure}[t]
\begin{center}
\includegraphics[height=2.2 cm]{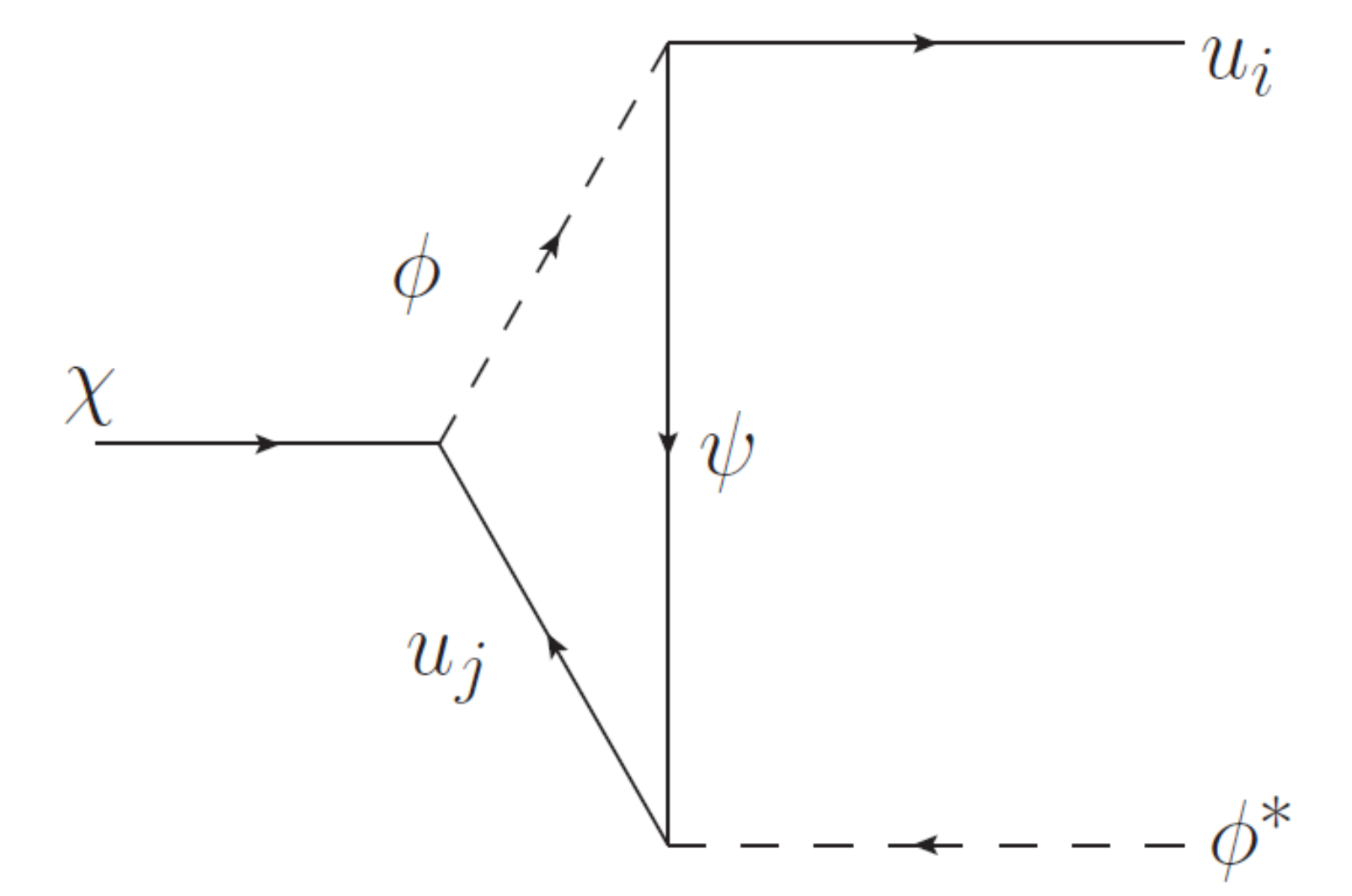}
\includegraphics[height=2.2 cm]{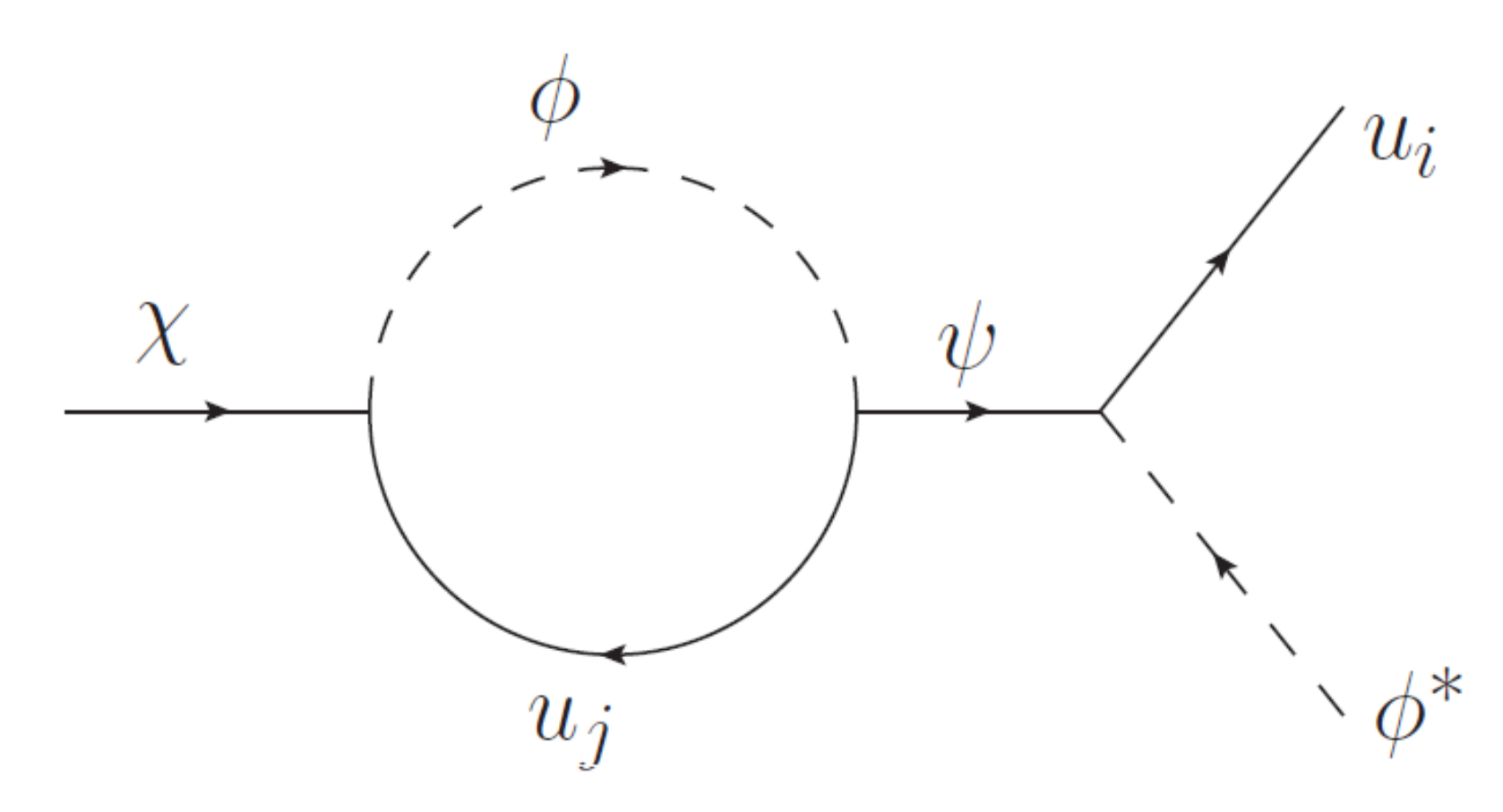}
\caption{Loop diagrams that interfere with tree-level decay to generate $\epsilon_{CP}$}
\label{fig:decaycpv}
\end{center}
\end{figure}

In addition to the asymmetry generating decay, there are other associated B-violating processes in this model that can potentially wash out the asymmetry. A detailed study of these washout effects can be found in \cite{Cui:2012jh}. Here we briefly summarize the main results. Early-time washout processes occurring at $T> \Lambda_{QCD}$ include inverse decay $ u d d\rightarrow \psi$ via an onshell ${\phi}^*$,  $2\rightarrow2$ scattering $\psi u\rightarrow\bar{d}\bar{d}$ via $\phi$-exchange, and $3\rightarrow3$ or $2\leftrightarrow4$ scattering of quarks to anti-quarks via $\phi,\psi$ exchange. The weak washout requirement can be satisfied as long as the WIMP decays after the freeze out of these washout processes around weak scale, which can be generally realized in this scenario as we discussed earlier. In such a weak washout regime, the washout processes are either suppressed by Boltzmann factor or suppressed by the positive high power-law $(T/M_{EW})^n$. At the later time after the QCD phase transition and then BBN, nucleons instead of quarks become new effective degrees of freedom. $n-\bar{n}$ oscillation is the relevant washout process to consider for these epochs. The weak washout condition requires that the $n-\bar{n}$ transition rate is slower than the Hubble expansion rate, which imposes constraints on model parameters through the effective $\cancel{B}$ Majorana mass $\delta m$ of neutron, as a result of the $uddudd$-type of operator. A much stronger limit on $\delta m$, $\delta m\leq 6\times10^{-33}\rm GeV\approx(10^{8}\rm sec)^{-1}$, comes from current day precision test in $n-\bar{n}$ oscillation reactor experiments \cite{Mohapatra:2009wp}. Nevertheless, it turns out that $n-\bar{n}$ oscillation does not give the strongest constraint on the model parameters $\lambda_{ij}$, because
in the minimal model, $\lambda_{ij}$ for $\phi d_i d_j$ has to be anti-symmetric in $i, j$. Consequently the $uddudd$ operator giving rise to $\delta m$ is highly suppressed \cite{Goity:1994dq}. More stringent constraint comes from $pp\rightarrow K^+K^+$ decay via higher dimensional $\cancel{B}$ operator, giving the bound $\lambda_{12}\lesssim10^{-7}$ for $m_\phi, m_\psi\sim1$ TeV, $y_i\sim1$\cite{Goity:1994dq}. In \cite{Cui:2012jh} constraints from other precision tests such as flavour-changing neutral current and neutron EDM are also discussed. The upshot is that, these current-day precision constraints require the new couplings to the first two generations of quarks to be suppressed. A simple way to satisfy these constraints is to consider a third-generation dominated pattern where the new fields couple mostly to $b, t$, with CKM-like suppressions to light quarks. With this choice, the washout processes in the early universe would be further strongly suppressed.

With the necessary third-generation dominated flavour pattern, this minimal model can be naturally embed in the ``natural SUSY'' \cite{Barbieri:1987fn} framework with $\cancel{B}$ R-parity violating (RPV) couplings\cite{Brust:2011tb}. Natural SUSY refers to a minimal realization of supersymmetry as a solution to the electroweak hierarchy problem, where the super-partners of top (stop) and bottom quarks (sbottom) have weak scale masses, while those of the first two generation quarks can be much heavier and decouple from the low energy spectrum. The experimental data from LHC Run-I at 8 TeV has imposed tight constraints on the masses of the first two generation squarks, and ``natural SUSY'' is a motivated scenario that generally remains viable, especially when combined with RPV decay of stop or sbottom. Now let us see how our minimal model maps to this SUSY framework. The minimal model we just discussed provides a blueprint for this mapping.
We simply promote singlets $\chi$ and $S$ to chiral superfields, and add them to the RPV MSSM.
Superpotential terms relevant to our setup are:
\bea
  \nonumber W&\supset& \lambda_{ij}TD_iD_j + \varepsilon' \chi H_uH_d + y_t QH_uT++\mu_\chi\chi^2\\
  &+& \mu H_uH_d+\mu_SS^2 + \alpha\chi^2S+ \beta SH_uH_d.\label{eq:susymodel}
\eea
We assume that the scalar component of $\chi$ is heavy and decouples, in the pattern similar to the first two generation squarks. The diquark $\phi$ in our minimal model is identified with the light $\tilde{t}_R$ in superfield $T$, Majorana $\psi$ is identified as a gaugino. The Lagrangian descendent from eq.(\ref{eq:susymodel}) enables the mixing between the SM Higgs and the singlet scalar $S$, as well as the mixing between $\chi$ and Higgsino. 
Consequently $\chi$ can annihilate through the Higgs portal, and decay into $\tilde{\bar{t}} t$ via $\chi-\tilde{H}_u$ mixing through the small coupling $\varepsilon' \chi$.\\

A further comment to make here is that, the intriguing scenario of natural SUSY with prompt RPV $\cancel{B}$ decay of stop as the smoking-gun signal at the LHC actually suffers from a potential cosmological crisis, for which our SUSY baryogenesis model provides a robust cure. Assuming a conventional baryogenesis generates baryon asymmetry at or above EW scale, such as through leptogenesis or EW sphalerons, the presence of RPV interactions ($\lambda_{ij}\gtrsim10^{-7}$) strong enough for prompt decays inside the LHC detectors would typically wash out these primordial B-asymmetry\cite{Barbier:2004ez, Brust:2011tb, Cui:2012jh}. Our model utilizes the RPV couplings that can erase any primordial asymmetry and then regenerate it through late decay that can occur below weak scale, after all the washout processes freeze out. There are alternative interesting solutions to this problem \cite{Dreiner:1992vm, Davidson:1996hs, Dimopoulos:1987rk, Cline:1990bw}, some also considered low scale baryogenesis in $\cancel{B}$ SUSY. But most of these mechanisms are less generic, or sensitive to details about cosmic initial conditions related to the inflaton or gravitino. In addition, these works do not address the WIMP miracle or $\Omega_{\rm DM}-\Omega_B$ ``coincidence''.

%%%%%%%%%%%%
\subsubsection{Bino Baryogenesis in Mini-split SUSY}
The last model example we will review is a neat incarnation of the WIMP baryogenesis idea in the mini-split SUSY framework, which is another general class of SUSY scenario that survives the existing LHC data. In the simplest version of mini-split SUSY \cite{ArkaniHamed:2012gw}, gaugino masses are of weak scale, and 1-loop factor suppressed compared to the sfermions masses. Although deviating from the conventional preference of perfect naturalness, mini-split SUSY is consistent with current constraints from flavour physics and LHC data in a general way (including the observed Higgs mass and limits on squark masses), while maintaining the merit of gauge coupling unification. The work in \cite{Cui:2013bta} pointed out that, by including RPV interactions, the \textit{minimal} spectrum/structure of the mini-split SUSY model (MSSM) provides all the ingredients for successful baryogenesis. In particular, bino serves as the metastable WIMP parent for baryons, with a long lifetime that naturally arises from the mass hierarchy between the sfermions and bino. It is interesting to note that in the R-parity conserving MSSM, a stable massive bino typically has relic abundance above the observed DM density due to a small annihilation cross section, and therefore is unsuitable to be a DM candidate. In contrast, in the baryogenesis mechanism we are considering, such would-be over-abundance of an unstable WIMP is necessary to compensate for the suppression factor $\epsilon_{CP}\frac{m_p}{m_\chi}$, and makes bino a desirable candidate for baryogenesis. A light wino or gluino can run in the loop diagram that interferes with the tree-level B-violating bino decay, and gives rise to CP asymmetry $\epsilon_{\rm CP}$. The automatic presence of these additional Majorana states in the MSSM is essential for generating a physical CP phase, as well as a new source of B-violation at loop-level and thus satisfy the requirement by the Weinberg-Nanopoulos theorem. 

The relevant Lagrangian terms for this bino-baryogenesis model are as follows: 
      \bea
      W&=&\mu H_uH_d+\frac{1}{2}\lambda^{ijk}L_iL_j\bar{e}_k+\lambda'^{ijk}L_iQ_j\bar{d}_k+\frac{1}{2}\lambda^{''ijk}\bar{u}_i\bar{d}_j\bar{d}_k+h.c.\\
      \mathcal{L}_{\rm gauge}&=&\frac{\sqrt{2}}{2}g_1(H_u^*\tilde{H}_u\tilde{B}-H_d^*\tilde{H}_d\tilde{B})+\sqrt{2}g_1Y_{{f}_{L/R,i}}\tilde{f}_i^{*L/R,\alpha}{f}_i^{L/R,\alpha}\tilde{B}\\\nonumber
&+&\sqrt{2}g_2\tilde{f}_i^{*L/R,\alpha}T^a{f}_i^{L/R,\alpha}\tilde{W}^a+\sqrt{2}g_3\tilde{f}_i^{*L/R,\alpha}T^a{f}_i^{L/R,\alpha}\tilde{g}^a+h.c.     \\
\mathcal{L}_{\rm soft}&=&-\frac{1}{2}M_1\tilde{B}\tilde{B}-\frac{1}{2}M_2\tilde{W}\tilde{W}-\frac{1}{2}M_3\tilde{g}\tilde{g}-\tilde{f}_i^{*L/R,\alpha}{\bf{(m_{L/R,\alpha}^2)}}_{ij}\tilde{f}_j^{L/R,\alpha}+h.c.,
       \eea
where $i, j$ are family indices, $\alpha$ labels a SM fermion species with certain gauge charges and $L/R$ indicates left-handed or right-handed. CP violation can arise from complex phases in $M_i$ and $\bf{(m_{L/R,\alpha}^2)}_{ij}$. Notice that although gaugino interactions are originally flavor diagonal in gauge basis, in split SUSY, flavor mixings from sfermion mass matrix $\bf{(m_{L/R,\alpha}^2)}_{ij}$ can be $O(1)$ with large CP phases, while still being consistent with experimental constraints. 

The paper \cite{Cui:2013bta} discussed two example models: a direct baryogenesis model with light gluino, and a leptogenesis model with light wino. Here we just review the leptogenesis as a representative, as the other model is in close analogy. The 3-body L-violating decay processes in this model are shown in Fig.\ref{fig:lg_decays}. There are competing B-conserving decays: $\tilde{B}\rightarrow L\bar{L}\tilde{W}$, $\tilde{B}\rightarrow H^*H\tilde{W}$, which are generally subleading or comparable in the parameter space of our interest.

\begin{figure}
  \centering{\includegraphics[width=40mm]{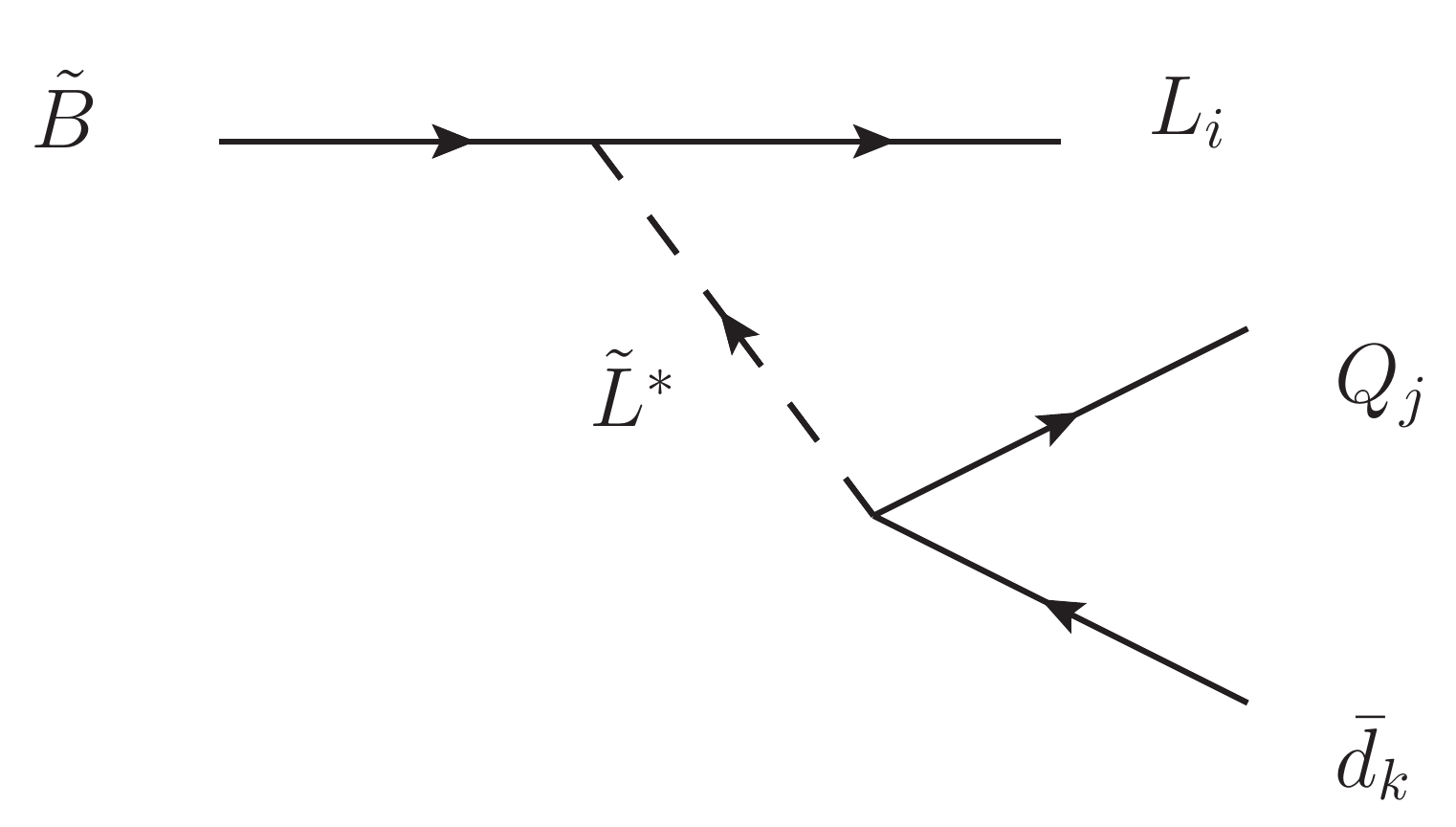}}  \\
  \begin{minipage}{5in}
        \includegraphics[width=50mm]{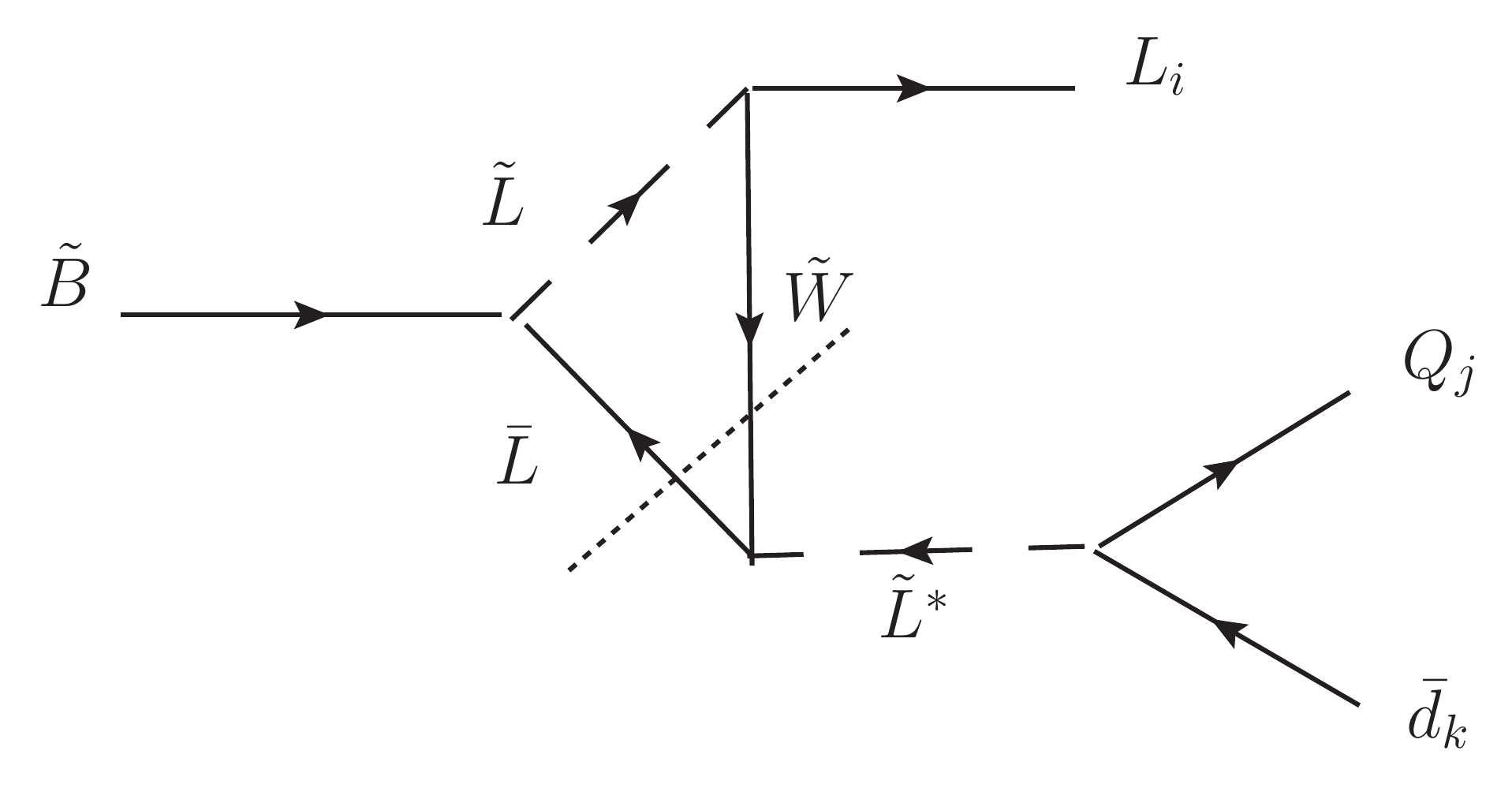} \qquad
         \includegraphics[width=50mm]{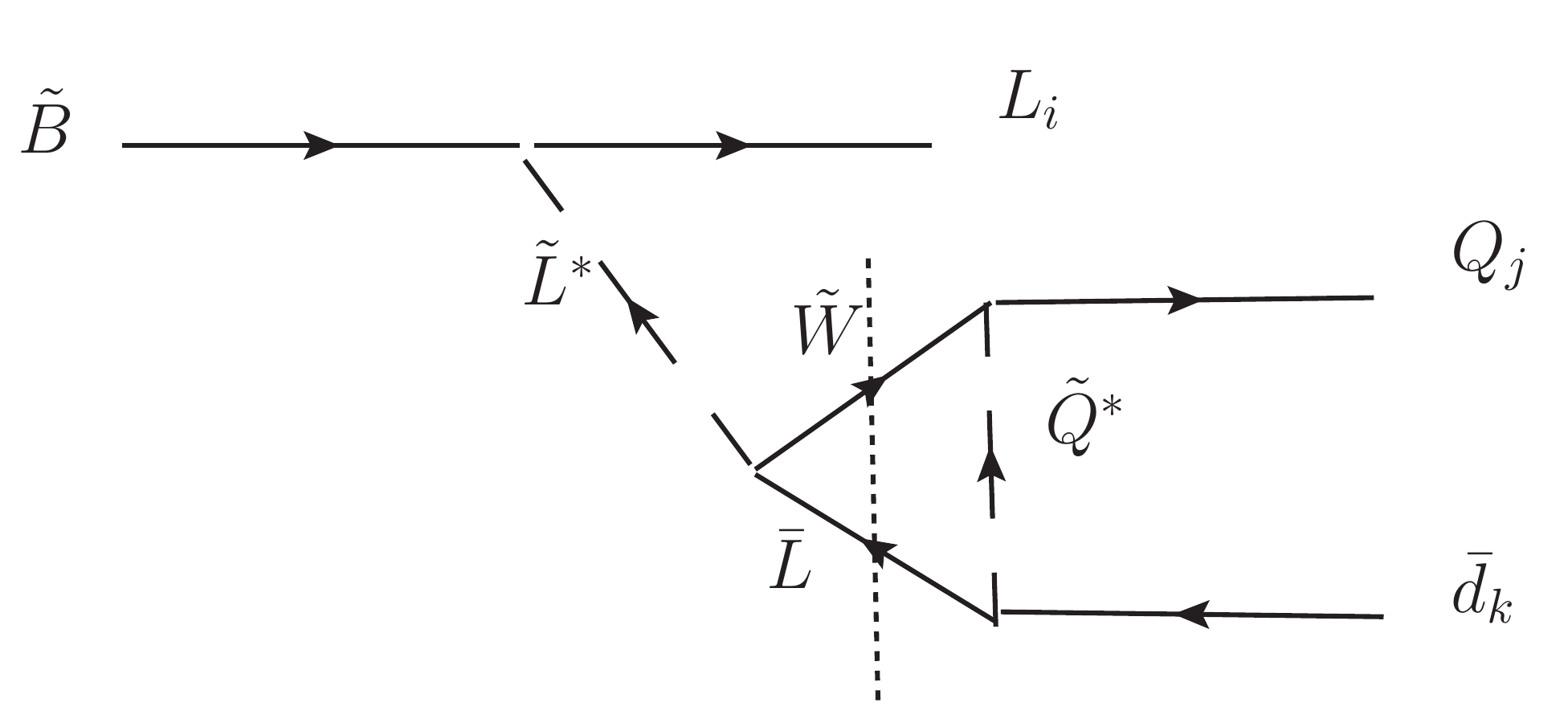} 
    \end{minipage}
  \caption{Decays of $\tilde{B}$ in the leptogenesis model. Upper: tree-level $\cancel{L}$ decay that triggers leptogenesis; \textit{Lower left}: produces a CP asymmetry by interference with tree-level decay even in the absence of flavor and CP violation in sfermion mass matrices.  \textit{Lower right}: contributes to CP asymmetry when the flavor and CP violation in sfermion mass matrices are sizable.}\label{fig:lg_decays}
\end{figure}

The CP asymmetry from the interference between tree-level $\cancel{B}$ decay and the lower left loop diagram in Fig.\ref{fig:lg_decays} is:
\be
 \epsilon_{CP}\equiv\frac{\Delta{\Gamma_{{\tilde{B},\cancel{B}}}}}{\Gamma_{{\tilde{B}}}}=\frac{g_3^2 Im[e^{i\phi}]C_2}{20\pi}\frac{m_{\tb}^2}{m_0^2},\label{epsiloncp_bg}
 \ee
 Here we see that the CP asymmetry is suppressed by the mass hierarchy between bino and the scalar superpartner mass $m_0$. A large would-be overabundance of bino is necessary to compensate such suppression, in order to obtain the observed $\Omega_B$.

Now the only missing piece for computing the asymmetric baryon abundance is the would-be relic abundance of bino, $\Omega^{\tau\rightarrow\infty}_{\tb}$. The leading annihilation channel is bino pair annihilation into Higgs pair via exchanging a Higgsino, as shown in Fig.\ref{fig:ann}. Other annihilation processes such as $\tb \tb\rightarrow d\bar{d}$ via squark exchange are more suppressed by the heavy mediator mass. 
\begin{figure}[t]
\begin{center}
\includegraphics[width=50mm]{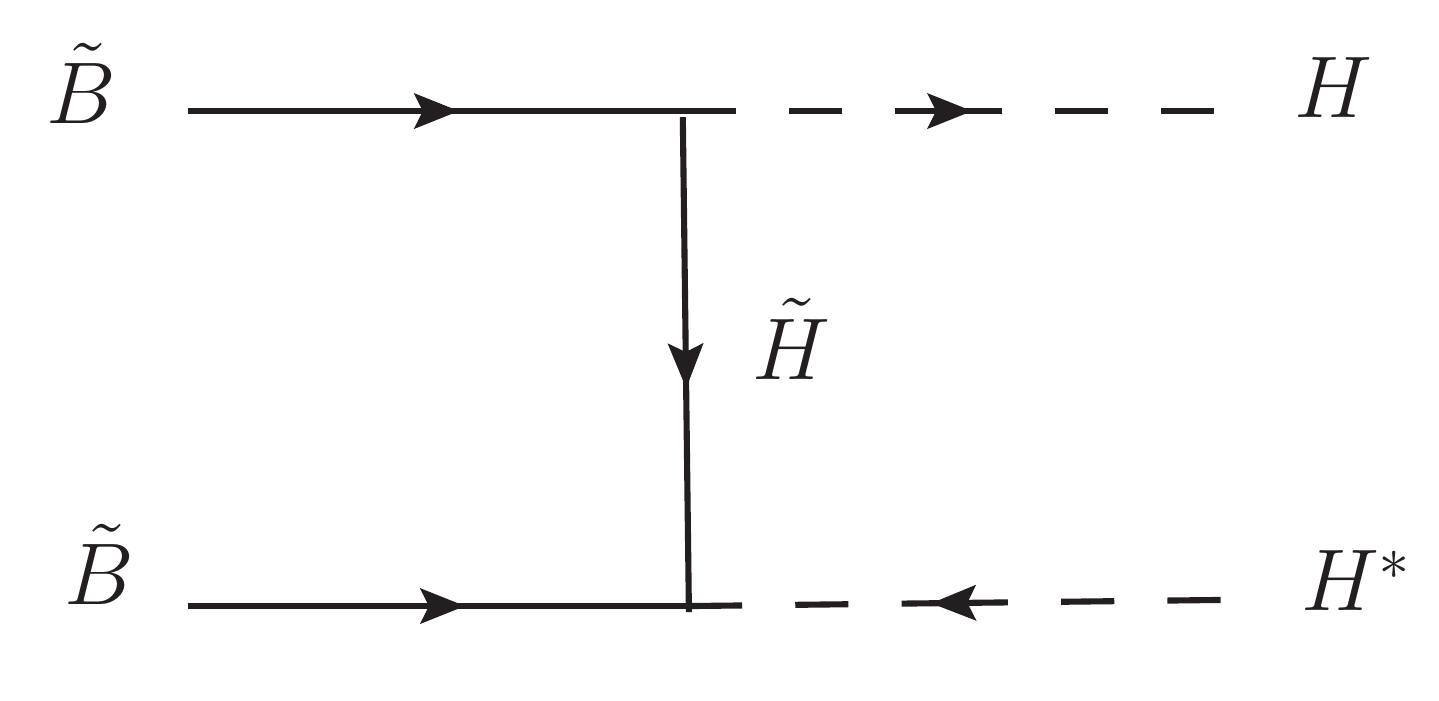}
\caption{Leading annihilation process of $\tb$}
\label{fig:ann}
\end{center}
\end{figure}

In order to have a small annihilation cross section and thus large would-be overabundance as desired, we need $\mu \gg m_{\tb}$. In such limit, the cross-section for the annihilation in Fig.\ref{fig:ann} is:
 \be
    \sigma_{HH^*}(s)=\frac{g_1^4}{32\pi} \frac{s-4M_1^2}{s\sqrt{1-4M_1^2/s}}\frac{1}{\mu^2}.
  \ee
  Taking the thermal average of $\sigma_{HH^*}(s)$ and using eq.(\ref{omegawimp}) we obtain an analytic estimate for the baryon abundance predicted in this model, assuming weak washout which can be generally realized:
  \be
  \Omega_{\Delta B}\sim10^{-2}\left(\frac{m_{\tb}}{1\rm~ TeV}\right)\left(\frac{\mu}{10m_0}\right)^2.\label{omegab_est}
\ee

The results from a numerical scan for sample benchmark points are shown in Fig.(\ref{fig:ssusy_results}), for both leptogenesis and direct bayrogenesis models. It is interesting to see that for a TeV bino, the viable parameter region for scalar superpartner mass $m_0$ determined purely by cosmological considerations happens to point to the mini-spilt regime of $10^2-10^4$ TeV. Fixing the mass ratios, while varying the overall mass scale away from the TeV scale, such bino baryogenesis mechanisms can still work. Nonetheless, TeV scale is about the lowest possible bino mass for a viable model in order to give sufficient baryon abundance (eq.\ref{omegab_est}), and to have its decay occurs at a temperature below its mass (weak-washout region). The criticality of a mini-split spectrum and the minimal bino mass at TeV scale for a successful baryogenesis in this scenario suggests that an imperfect naturalness in SUSY may result from a compromise between naturalness principle and cosmological environmental selection. This can be seen as an example in analogy to the ``galactic principle'' or ``atomic principle'' discussed in \cite{ArkaniHamed:2004fb,Giudice:2006sn,Weinberg:1987dv,Agrawal:1997gf}.

\begin{figure}
        \includegraphics[width=60mm]{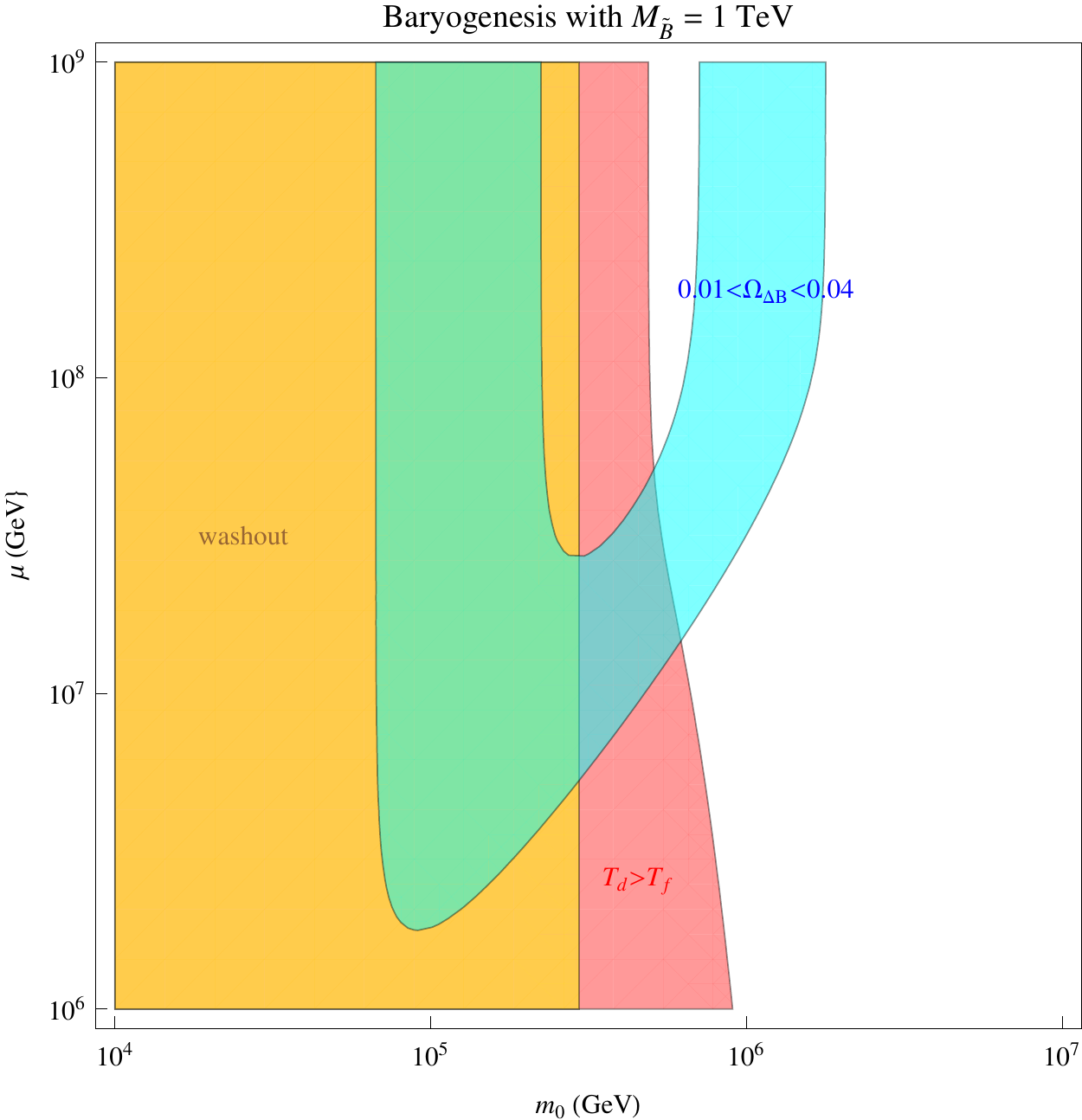} 
    \includegraphics[width=60mm]{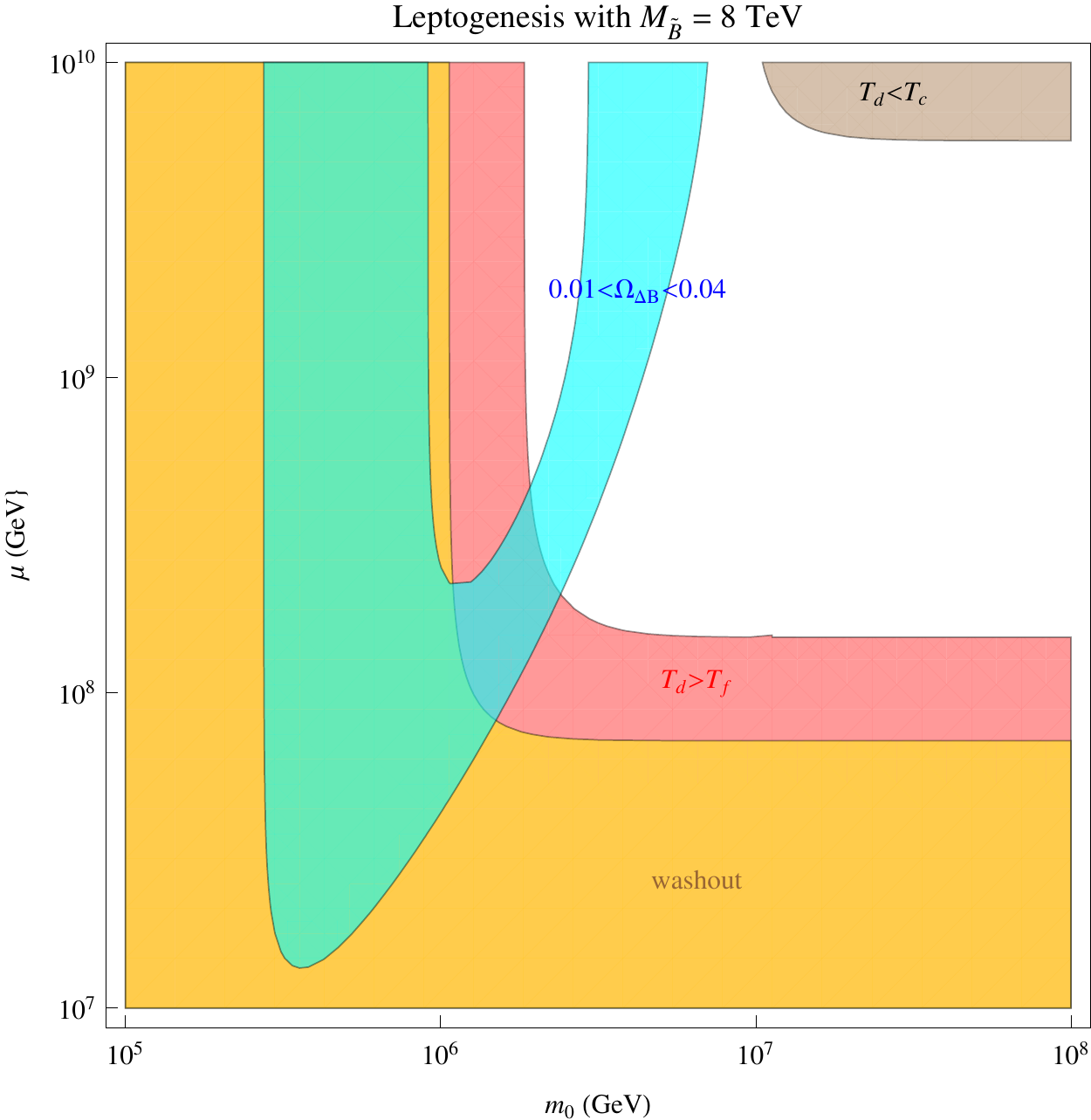}  
 \caption{Cosmologically allowed regions of parameter space for (left) baryogenesis and (right) leptogenesis models. We set RPV couplings $\lambda^{''}=\lambda^{'}=0.2$, $\phi=\frac{\pi}{2}$. Cyan region provides baryon abundance $10^{-2}<\Omega_{\Delta B}<4\cdot10^{-2}$. In the case of leptogenesis the brown region is excluded by decay after EWPT at $T_c\approx100\rm ~GeV$. The pink region is excluded by our simple basic assumption that bino decays after freezeout. The yellow region is excluded by requiring that washout processes are suppressed ($T_d<M_{\tb}$). The yellow region is in fact all included in the pink region (so it appears to be orange in the overlapping region). } \label{fig:ssusy_results}
\end{figure}
  %%%%%%
\subsection{Phenomenology}
Just like WIMPy baryogenesis from WIMP DM annihilation, the mechanism of baryogenesis from metastable WIMP decay can also be tested in a variety of experiments.
\bit
\item Indirect test: Intensity frontier experiments. The CP-violation and B-violation effects essential for these baryogenesis models can be searched for through precision measurements tests such as in $n-\bar{n}$ oscillation experiments and neutron EDM measurements. The strength of this type of signal is model-dependent. As discussed in \cite{Cui:2012jh, Cui:2013bta}, these models can be subject to the constraints from current limits by these precision tests, yet with generic parameter space that is allowed. This also implies that with the expected significant improvement in the coming years, these intensity frontier experiments, in particular the EDM measurements, can be sensitive to certain models/parameter space of this new baryogenesis mechanism. This is a direction that is worth further investigation.\\

\item Direct test: Collider experiments (LHC). The WIMP baryogenesis mechanism discussed in this section has a quite intriguing prediction for its collider signal, which is almost model independent. Recall that in order to satisfy the Sahakrov out-of-equilibrium condition for baryogenesis in a robust way (i.e. decay in the weak washout regime), the meta-stable WIMP $\chi$ needs to survive its thermal freeze-out time, which is around the weak scale. This imposes a cosmological condition for the proper lifetime of the WIMP:
\be
\tau_\chi\gtrsim t_{\rm fo}=0.3g_*^{-1/2}\frac{M_{\rm pl}}{T_{\rm fo}^2}\sim \left(\frac{T_{\rm fo}}{\rm 100 \,\,GeV}\right)^{-2}10^{-10}\rm \,\,sec,
\ee
which corresponds to a lower limit on its proper decay length of $\sim 1$ mm. Intuitively a lower limit of $\sim$1 mm corresponds to the size of our universe around the electroweak phase transition. The remarkable coincidence is that such macroscopic decay lengths of $\sim$1 mm corresponds to the tracking resolution of the detectors at collider experiments. Therefore, once being produced, the WIMP baryon parent would generate a displaced vertex within the detector of collider experiments such as the LHC, or the WIMP may escape the detector entirely, revealing itself as missing energy. The former channel of displaced vertex signal is particularly intriguing as the search for long-lived new particles in this channel typically bear very low background from SM events, and thus can be sensitive to very rare new physics signal events. Nonetheless, such searches were not well developed, in contrast to that for prompt decay or missing energy, yet have attracted rising interest and endeavour recently from both theorists and experimentalists. This new WIMP baryogenesis idea that we proposed provides a cosmological motivation for further developing the displaced vertex searches, and can serve as a signal generator to test the experimental coverage for a variety of final state possibilities. In fact such cosmological motivation for displaced vertex at colliders can generally arise from baryogenesis triggered by weak scale massive particle decay, with WIMP baryogenesis as a particularly motivated example. In \cite{Cui:2014twa} a simplified model approach was employed to facilitate the study of collider phenomenology for these models. The production channels include electroweak process and Higgs portal (through on-shell or off-shell SM higgs decay, or heavy Higgs resonance decay) and the decay final states can be classified by RPV interactions. Due to the approximate $Z_2$ symmetry that is generally associated with these models to ensure a long-life time of the WIMP, the WIMPs are typically pair produced. The topology of such search for metastable WIMPs are quite general, and is in analogy to the missing energy search for WIMP DM. We illustrate both cases in Fig.\ref{fig:WIMP_schematic} for direct comparison. Also in \cite{Cui:2014twa} a recast of 8 TeV LHC data from CMS displaced dijet search was performed  to give constraints on model parameter space, and the projection of sensitivity reach at 14 TeV was made with suggestions for improvements. \\

An updated note to add here is that, the work in \cite{Cui:2014twa} has drawn significant interest from the displaced hadronic jets working group in the ATLAS experiment collaboration, and the simplified models of WIMP baryogenesis will be included as a benchmark examples in the official analysis based on the coming 14 TeV data.
A more ambitious yet challenging goal is to measure the CP violation and even B- or L-number violation in such WIMP baryon parent decays. However, this would require a dedicated search with high luminosity, and probably next generation collider experiments.
\eit

\begin{figure}
\begin{center}
{\includegraphics[width=0.4\textwidth]{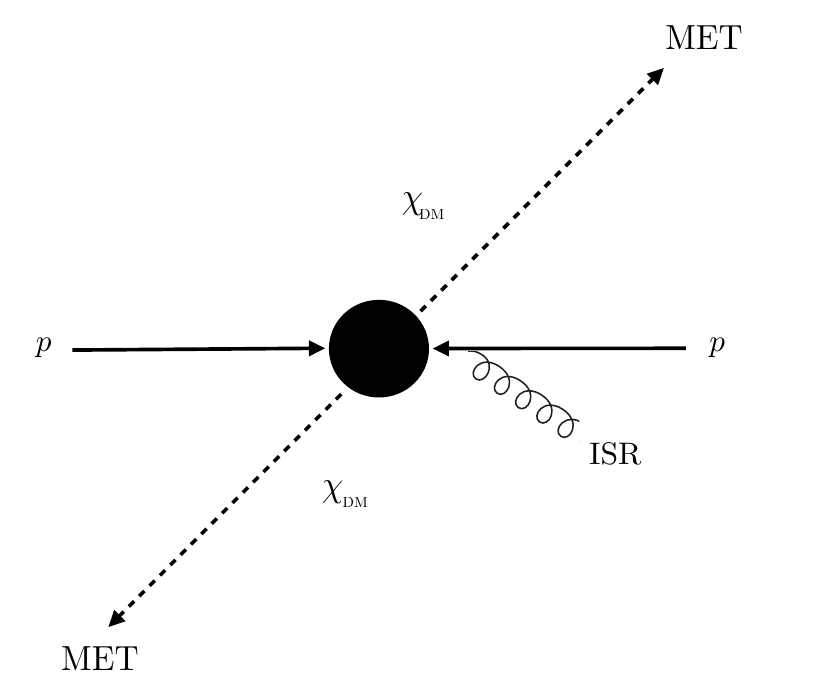}}
{\includegraphics[width=0.4\textwidth]{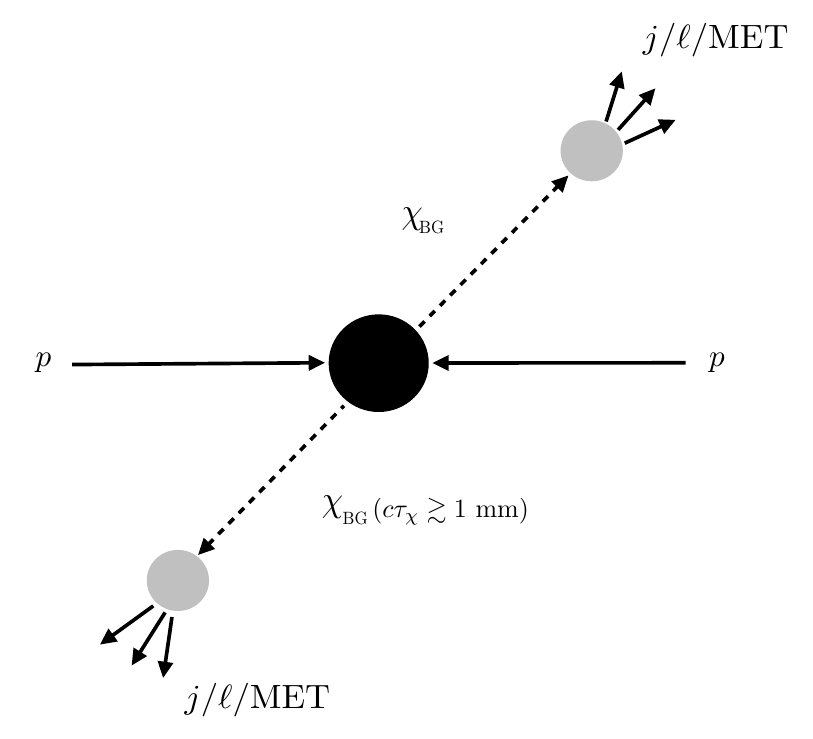}}
\caption{Schematic diagram showing the pair-production at the LHC of: (left) dark matter in stable WIMP dark matter searches, with associated initial state radiation (ISR); (right) the analogous production of the meta-stable WIMP triggering baryogenesis, which decays at a displaced vertex to jets, leptons and/or missing transverse energy.}
\label{fig:WIMP_schematic}
\end{center}
\end{figure}

\textbf{Summary:}
\bit
\item Baryogenesis from metastable WIMP decay is a new mechanism to generate baryon asymmetry around or below weak scale cosmic temperature. It has a WIMP-miracle type of prediction for baryon abundance that is insensitive to details about the washout effects, as the weak washout regime can be generally realized with late decay.
\item This baryogenesis framework does not have dark matter as a built-in ingredient, so it is compatible with non-WIMP type DM. Nonetheless, it is straightforward to incorporate another stable WIMP as DM, in which case the mechanism provides a novel path to address the cosmic coincidence between DM and baryon abundances. It is also possible that a tighter connection to DM can be realized with further model building effort.

\item This baryogenesis mechanism has model-dependent implications for intensity frontier experiments such as neutrino EDM, $n-\bar{n}$ oscillation, which is worth further study. More remarkably, it has model-independent implications for collider experiments, in particular it motivates displaced vertex searches from cosmological consideration.\eit
\section{Conclusions}
In this review  letter we summarized two recently proposed baryogenesis mechanisms triggered by the WIMP-type of new particles: baryogenesis via out-of-equilibrium annihilation of WIMP DM during thermal freeze-out, and baryogenesis via a metastable WIMP decay after thermal-freezeout. These new mechanisms can address the cosmic abundances of dark matter and atomic matter, as well as the coincidence between the two. Within these new frameworks, there is great potential for further theoretical development in the direction of model-building. In addition, with new particles and interactions at the electroweak scale, these WIMP baryogenesis mechanisms offer exciting opportunities for probing the cosmic origin of our atomic matter with current-day experiments, in close analogy to the prospect of detecting WIMP dark matter. In particular, baryogenesis from long-lived WIMP decay is now being actively pursued by the displaced vertex searches with the coming data at the Large Hadron Collider. WIMP baryogenesis mechanisms provide new examples of the fascinating possibility that new particle physics related to the electroweak hierarchy problem may also address important puzzles in modern cosmology, and is worth further exploration in both theoretical and phenomenological aspects.

%\appendix

%\section{Appendices}

\section*{Acknowledgments}

The author is supported by Perimeter Institute for Theoretical Physics, which is supported by the 
Government of Canada through Industry Canada and by the Province of 
Ontario through the Ministry of Research and Innovation. The author is also supported in part by the US NSF under grant PHY-1315155 and the Maryland Center for Fundamental Physics. 
%\section*{References}

\end{document}